\documentclass[useAMS,usenatbib]{mn2e}
\usepackage{graphicx,rotating,subfigure,floatflt,amsmath,wasysym}
\usepackage{here}
\title{Determining the extragalactic extinction law with SALT}

\author[Ido Finkelman et al.]{Ido Finkelman$^{1}$\thanks{E-mail:
ido@wise.tau.ac.il (IF); noah@wise.tau.ac.il (NB); akniazev@saao.ac.za (AYK); dibnod@saao.as.za (DB; dod@saao.ac.za (DOD); hashimot@saao.ac.za (YH); nsl@saao.ac.za (NL); erc@saao.ac.za (ER); still@saao.ac.za (MS); petri@saao.ac.za (PV)}, Noah Brosch$^{1}$, Alexei Y. Kniazev$^{2}$, David Buckley$^{2}$,
\newauthor Darragh O'Donoghue$^{2}$, Yas Hashimoto$^{2}$, Nicola Loaring$^{2}$, Encarni Romero$^{2}$, 
\newauthor Martin Still$^{2}$, Petri Vaisanen$^{2}$\\
$^{1}$The Wise Observatory and the Raymond and  Beverly Sackler School of Physics and
Astronomy, the Faculty of Exact
Sciences, \\ Tel Aviv University, Tel Aviv 69978, Israel\\
$^{2}$SALT, South African Astronomical Observatory, P.O. Box 9,
Observatory 7935, Cape Town, South Africa}

\begin{document}

\pagerange{\pageref{firstpage}--\pageref{lastpage}} \pubyear{2002}

\maketitle

\label{firstpage}

\begin{abstract}
We present CCD imaging observations of early-type galaxies with
dark lanes obtained with the Southern African Large Telescope
(SALT) during its performance-verification phase. The observations
were performed in six spectral bands that span the spectral range
from the near-ultraviolet atmospheric cutoff to the near-infrared.
We derive the extinction law by the extragalactic dust in the dark
lanes in the spectral range $ 1.11 \mu m^{-1} <\lambda^{-1}<2.94 \mu m^{-1}$ by fitting model galaxies to the unextinguished parts of the
image, and subtracting from these the actual images. This procedure
allows the derivation, with reasonably high signal-to-noise, of
the extinction in each spectral band we used for each resolution element of the image. We also introduce an alternative method to derive the extinction values by comparing various colour-indices maps under the assumption of negligible intrinsic colour gradients in these galaxies. We than compare the results obtained using these two methods.

We compare the total-to-selective extinction derived for these
galaxies with previously obtained results and with similar
extinction values of Milky Way dust to derive conclusions about the
properties of extragalactic dust in different objects and
conditions.

We find that the extinction curves run parallel to the Galactic extinction curve, 
which implies that the properties of dust in the extragalactic enviroment are 
similar to those of the Milky Way, despite our original expectations. The ratio of the total V band extinction 
to the selective extinction between the V and B bands is derived for each galaxy with an 
average of $2.82\pm0.38$, compared to a canonical value of 3.1 for the Milky Way. The similar values imply that galaxies with well-defined dark lanes have characteristic dust grain sizes similar to those of Galactic dust. We use total optical extinction values to estimate the dust mass for each galaxy, compare these with dust masses derived from IRAS measurements, and find them in the range $10^4$ to $10^7M_{\odot}$.  
\end{abstract}

\begin{keywords}
galaxies: early type, dust: extinction
\end{keywords}

\section*{Introduction}
Since its existence was first inferred from the obscuration of starlight (Trumpler 1930), interstellar dust has been  recognized as an important component of the interstellar medium (ISM) playing role in the evolution of galaxies, the formation of stars and planetary systems, and possibly, the origins of life.
Studying the properties of the dust particles in interstellar space can help understand the nature of the dust grains themselves, the processes that govern their evolution and the way those grains affect the light of stars in their vicinity. 
The last property is mainly due to the interaction of dust grains with the electromagnetic radiation, i.e., due to the scattering and absorption of light, which depend strongly on the size distribution, structure and chemical composition of the grains. This attenuation of starlight is wavelength-dependent and is expressed as the wavelength dependence of the extinction, or the ``extinction law''. Since extinction tends to be greater in blue than in red, this is often referred to as ``reddening''. 

The extinction law in the Milky Way (MW) has been studied extensively for many decades since it was first measured for the optical region by Rudnick (1936) using the still-widely-used ``pair match''. This method is based on the comparison of the spectral energy distribution (SED) of pairs of stars of exactly the same spectral and luminosity classes. One of the stars is usually selected to be nearby thus is assumed to be clear of dust with a very small chance of being significantly extinguished. Any difference between the two SEDs is attributed to dust extinction and the global extinction law is derived after comparing many pairs of stars. From the far-UV to the far-IR, the standard MW extinction law is generally adopted to be that derived by Savage \& Mathis (1979; hereinafter S\&M). However, there are significant differences between different lines of sight that cause large variations. 
The S\&M law was revised by Cardelli et al.\ (1988, 1989) and $R_V$, the ratio of total to selective extinction, was proposed as the single-parameter value determining it and varying in different environments. While $R_V$ is typically 3.1, values as low as 2.1 and as high as 5.6 have been measured (Valencic et al.\ 2004). An explanation of the general extinction law as a manifestation of the size distribution of dust grains was proposed by Greenberg \& Chlewicky (1983, see their Fig.\ 1). Measuring the wavelength dependence of the extinction offers, therefore, a means of estimating the size of dust grains in the MW and in other galaxies.

The extinction law in other galaxies has been measured using the pair method only for relatively nearby objects, such as the Large and Small Magellanic Clouds where individual stars can be resolved (e.g., Gordon et al.\ 2003), and significant differences from the MW extinction law were observed. As far as other galaxies are concerened, the above method is usually not practical due to the difficulty to resolve individual stars, thus other methods must be used.
Some of the methods suggested in the literature includes measuring the extinction by foreground galaxies of distant quasars (Ostman, Goobar \& Mortsell 2008), determining the differential extinction of multiply-imaged quasars (Falco et al.\ 1999), determining the dust extinction of gamma-ray burst host galaxies (Li, Li \& Wei 2007) and measuring the extinction law in starburst galaxies (Calzetti et al.\ 2000). However, in this article we adopt a different method, which studies the extinction law in early-type galaxies with dust lanes. 

The existence of early-type galaxies with dust lanes was recognized following the identification and study of five ellipticals with minor-axis dust lanes by Bertola \& Galleta (1978), and the publications of galaxy catalogs for objects with dust lanes by Hawarden et al.\ (1981), Ebneter \& Balick (1985) and V\'{e}ron-Cetty \& V\'{e}ron (1988).
Moreover, many early-type galaxies show the presence of a multiphase ISM with interstellar dust as an important constituent. This is indicated by recent surveys revealing the presence of dust extinction in a large fraction of early-type galaxies (Van Dokkum \& Franx 1995; Ferrari 1999; Tomita et al.\ 2000; Tran et al.\ 2001). 
The method is based on the assumption that the underlying galaxy, whether elliptical or lenticular, has a smoothly-varying light distribution while the dust lanes are only local disturbances of the global brightness pattern. If the dust covering factor of the galaxy is relatively small, one can extract the dust-free model of the original underlying galaxy. Comparing the light distribution of the unextinguished galaxy with the actual observed one allows the derivation of extinction at each measured angular resolution element of the target galaxy. This method was used for the first time to derive the properties of dust in the lenticular galaxy N7070A (Brosch et al. 1985). A discussion of the extragalactic extinction law in early-type galaxies, following a number of other studies, was summarized by Brosch (1988). Later attempts to derive extragalactic extinction law in early-type galaxies were by Brosch et al.\ (1990), Brosch and Loinger (1991) Goudfrooij et al.\ (1994), Sahu, Padney \& Kembhavi (1998), Patil et al.\ (2002) and Patil et al.\ (2007). 

Brosch \& Loinger (1991) used a simple elliptical-isophote fit to model the unextinguished part of the galaxy NGC 7625 and derived an extinction law rising faster at its blue end than the accepted MW standard.
In latter publications by Goudfrooij et al.\ (1994), Sahu et al.\ (1998), Patil et al.\ (2002) and Patil et al.\ (2007), the light distribution of the underlying galaxy was modeled after first masking off the dust lanes. This was done first for the reddest spectral band (usually the I band) by allowing a free-parameter fit to determine the optical center of the galaxy in a band that is least obscured by dust. This fit was repeated for the bluer bands keeping the galactic center fixed. This way, the extinction at a specific location in the galaxy was determined for each wavelength used in the observations. The overall extinction laws were compared with the standard MW extinction law.

One overall result found in most studies of extragalactic dust is its similarity to the dust in the MW, at least in the properties revealed in the optical part of the spectrum. This is amazing, considering the wide variety of galaxies and, presumably, of physical conditions of the dust. Unless some cosmic plot is afoot, one would expect dust grains of different sizes and different chemical compositions in different objects - yet this does not appear to be the case.

While a significant number of early-type galaxies have been investigated with this method (e.g., 26 objects in Patil et al.\ 2007 and 10 objects in Goudfrooij et al.\ 1994, none in common), in most cases the measurements were restricted to bands longwards of B. As noted by Nandy (1984) from studies of the Magellanic Clouds extinction, differences among extinction curves are better detected at shorter wavelength such as the U-band or the space ultraviolet. This paper aims to extend the optical study of the extinction law in a sample of galaxies from the standard B, V, R \& I bands to the near-UV by using two filters that split the U-band into two intermediate spectral regions. This allows the extraction of more information about the extragalactic extinction law close to the blue atmospheric cutoff. The observations were performed with a very large telescope, which allowed a reasonable sensitivity even at the shortest spectral band used, despite the atmospheric extinction.

This paper is organized as follows: \S~\ref{txt:Obs_and_Red}
gives a description of all the observations and data reduction. In
\S~\ref{txt:results} we present our results, analyze them in
\S~\ref{txt:analyse}, and present our interpretation in
\S~\ref{txt:interp}. We discuss the results in \S~\ref{txt:disc} and summarize our conclusions in \S~\ref{txt:summ}.
\begin{table*}
 \centering
 \begin{minipage}{110mm}
  \caption{Global parameters for galaxies in our sample.
  \label{t:Obs}}
\begin{tabular}{lcclccc}
\hline
Object     & RA       & DEC       & Morph.\       & B$^0_T$ & v$_{Helio}$  & Size \\
{}         & (J2000.0)& (J2000.0) & (RC3)         &   {}    &   (km/s)     & (arcmin)\\
\hline 
NGC3283    & 10:31:11 & -46:15:02 & S0            &  12.50  &    2889      & 2.6x1.5\\
NGC3497    & 11:07:18 & -19:28:21 & S0            &  13.03  &    3707      & 2.6x1.4\\
AM1118-290 & 11:20:53 & -29:24:09 & S0            &  14.22  &    8197      & 1.1x0.7\\
NGC4370    & 12:24:55 & +07:26:38 & Sa            &  13.48  &     722      & 1.4x0.7\\
AM1352-333 & 13:55:34 & -33:54:05 & S0            &  14.30  &    4577      & 1.1x0.9\\
NGC5626    & 14:29:49 & -29:44:56 & SA0+          &  13.98  &    6853      & 1.2x0.9\\
AM1459-722 & 15:04:35 & -72:37:00 & S?            &  14.05  &    4493      & 1.3x0.9\\
NGC5799    & 15:05:36 & -72:25:57 & S0            &  13.32  &    3099      & 1.3x1.0\\
NGC5898    & 15:18:13 & -24:05:49 & E0            &  12.49  &    2209      & 2.2x2.0\\
\hline
\end{tabular}
\end{minipage}
\end{table*}
\begin{table*}
 \centering
 \begin{minipage}{170mm}
\caption{Observing log. \label{t:exposure}}
\begin{tabular}{|l|lllllll|cccccc|}
\hline
Galaxy & \multicolumn{7}{|c|}{Exposure time (sec)} & \multicolumn{6}{|c|}{point source FWHM (")}\\
{}         & Date       & B        & V        & R        & I        & U2      & U1      & B    & V    & R    & I    & U2  & U1 \\
\hline
NGC3283    & 16.03.2007 & $20(7)$  & $20(7)$  & $ 10(7)$ & $10(7)$  & $120(7)$  & $120(7)$ & 1.82 & 1.74 & 1.84 &   1.82 & 2.09 & 1.85\\
NGC3497    & 24.02.2007 & $30(6)$  & $30(6)$  & $ 20(5)$ & $20(5)$  & $120(5)$  & $120(4)$ & 2.38 & 2.30 & 2.41 &   2.18 & 2.41 & 2.38\\
AM1118-290 & 26.02.2007 & $30(5)$  & $30 (5)$ & $ 20(5)$ & $20(5)$  & $120(4)$  & $120(4)$ & 2.02 & 1.74 & 1.59 & 1.67 & 1.92 & 2.11\\
NGC4370    & 23.02.2007 & $30(8)$  & $30(6)$  & $ 20(8)$ & $20(8)$  & $120(8)$  & $120(14)$ & 2.27 & 2.37 & 2.04 &   2.13 & 2.35 & 2.35\\
AM1352-333 & 26.02.2007 & $30(5)$  & $30 (5)$ & $ 20(5)$ & $20(5)$  & $120(4)$  & $120(9)$ & 1.96 & 1.88 & 1.88 & 1.85 & 1.79 & 2.07\\
NGC5626    & 17.05.2007 & $30(9)$  & $30(10)$ & $ 20(8)$ & $20(7)$  & $120(6)$  & $120(6)$ & 2.49 & 2.41 & 2.27 & 2.60 & 2.55 & 2.66\\
AM1459-722 & 16.03.2007 & $30(9)$  & $30(10)$ & $ 20(9)$ & $20(10)$ & $120(10)$ & $120(9)$ & 1.90 & 1.96 & 1.88 & 1.90 & 1.51 & 1.57\\
NGC5799    & 23.02.2007 & $20(8)$  & $20(8)$  & $ 10(8)$ & $10(8)$  & $120(8)$  & $120(8)$ & 2.32 & 2.27 & 2.21 & 2.18 & 2.24 & 2.60\\
NGC5898    & 21.02.2007 & $6(6)$   & $18(6)$  & $ 25(6)$ & $45(6)$  & $76(6)$   & $92(6)$ & 1.88 & 1.9 & 1.65 & 1.73 & 1.73 & 1.89\\
\hline
\end{tabular}
\begin{minipage}[]{6.5in}
\begin{footnotesize}
{\bf Note:} The number in parenthesis represents the number of individual frames combined to produce the final galaxy image in each band. The point source FWHM represents the typical FWHM of stars near each target combining seeing and image quality effects.
\end{footnotesize}
\end{minipage}
\end{minipage}
\end{table*}

\section{Observations and data reduction}
\label{txt:Obs_and_Red}
The Southern African Large Telescope (SALT) was described by
Buckley, Swart \& Meiring (2006), and its CCD
camera (SALTICAM) was described by O'Donoghue et al.\ (2006).
We used the SALT and SALTICAM to
observe dark-lane early-type galaxies.

It may seem superfluous to observe bright galaxies with an
(effectively) nine-meter telescope since smaller telescopes served admirably this purpose in the past. These observations, however, were obtained during the Performance Verification (PV) phase of
the SALT telescope when only the SALTICAM 
was available, and even this with the instrument before its
upgrade to auto-guiding. The telescope was still affected by the
limited image quality, which restricted the size of the workable
field to some three arcmin instead of the nominal eight arcmin region. The specific advantage in performing
this project with SALT and SALTICAM is the high efficiency of this
combination in the near-ultraviolet, and the availability of the
U1 and U2 filters (see below) among the SALTICAM filter
complement.
\begin{figure}
\includegraphics[width=84mm]{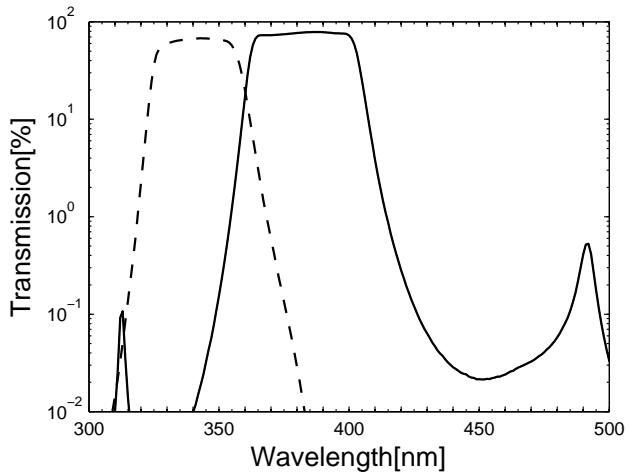}
\caption{SALTICAM short wavelength filters: 340/45 nm (U1- dashed line) and 380/40 nm (U2- solid line) transmission.}
\label{fig:transmission}
\end{figure}

The SALTICAM is a CCD mosaic of two 2048x4102 pixels CCDs and therefore, for consistency, program objects were imaged with the same CCD throughout the observations. The full SALTICAM image is about 9'.6 by 9'.6 with pixels of $\sim$0".28 (after binning on-chip by a factor of two), but the nominal science field is eight arcmin.

Observations of nine dust lane early-type galaxies were performed in service mode from February to May 2007 with SALT, whereas only for eight objects we derived the extinction values (see \S~\ref{txt:results}). The nine galaxies are listed in Table \ref{t:Obs} with their coordinates, morphological classification and optical size as taken from the RC3 catalog (de Vaucouleurs et al.\ 1992) or from LEDA (http://leda.univ-lyon1.fr).
CCD imaging observations were performed with SALTICAM using the standard B, V, R \& I filters, as well as two short-wavelength interference filters, U1 and U2, with transmission bands peaking at 340 nm (FWHM 35 nm) and 380 nm (FWHM 40 nm) respectively, as shown in Fig.\ \ref{fig:transmission}.
The plotted transmission profile is
logarithmic, to emphasize the reasonably good blocking of
out-of-band wavelengths.
U1 has its strongest out-of-band
transmission band above 1 $\mu$m where the SALTICAM CCD already
has a fairly low quantum efficiency. U2 is slightly more affected
by a red ``leak'' given its prominent transmissions at $\sim$500 nm and
at $\sim$700 nm. However, the first out-of-band peak at 480 nm is at less than $1\%$
level, and the second one, a blue leak near 310nm,
is at $\sim 0.1\%$ level.

The objects were selected from catalogs for objects with dust lanes by Hawarden et al.\ (1981), Ebneter \& Balick (1985) and V\'{e}ron-Cetty \& V\'{e}ron (1988). The observed objects were chosen according to visibility conditions and availability of observing nights requiring that the galaxies would fit in the good image quality field of SALTICAM ($\sim$ three arcmin), would be in the sky region accessible to SALT at the date of observation, and would not present reduction complications such as dust lanes with bright edges that might signify internally-produced light within the dust lanes. Galaxies from Patil et al.\
(2007) and from Goudfrooij et al.\ (1994) need the full observation set even though they were already
observed by these authors in some of the bands. This is because the images to be
used for the derivation of the extinction should all have the same
seeing size. This is ensured by obtaining the six images
quasi-simultaneously, with the same camera and telescope.
The observations consist of short-duration integrations while the filter wheel cycles between the six filters, thus ensuring quasi-similiar seeing size for all bands observed in a single filter cycle.
Exposure times are determined to be sufficiently short to prevent image trailing when operating in the SALTICAM unguided mode (but with SALT tracking) or overexposure of the brightest part of each galaxy. The six-image cycles were repeated while the pointing was slightly shifted to allow for a dithering effect. The cycles were then accumulated to
reach better exposure depth. 
Exposure times in seconds for B, V, R, I, U2 \& U1 bands, as well as the typical FWHM of stars near target in the different bands, are listed in Table \ref{t:exposure}. The numbers in parenthesis represent the number of exposures acquired in a given band. Since each image consists of a combination of a number of short-duration exposures, the FWHM value represents the value to which the different images in each band were convolved. The size of the resolution elements is a result of both seeing and image quality (IQ) effects. 

The SALT pipeline includes bias subtraction, overscan subtraction and ``cross-talk'' correction, while other standard preprocessing steps were done using standard tasks within IRAF\footnote{IRAF is distributed by the National Optical Astronomy Observatories (NOAO), which is operated by the Association of Universities, Inc. (AURA) under co-operative agreement with the National Science Foundation}. Such steps include geometric alignment of multiple frames taken in each filter by measuring centeroids of several common stars in the galaxy frames. The frames were then combined with median scaling to improve the S/N ratio. This alignment procedure involves IRAF tasks for scaling, translation and rotation of the images, so that a small amount of smearing is introduced affecting the accuracy to be better than half a pixel.
Median combination was also useful in removing cosmic rays events while the CCD hot pixels were removed with the CCDMASK task in IRAF using an appropriate mask. We emphasize that no flatfield correction was made during this  reduction process since at the time of observations the telescope did not yet have a moving baffle to simulate the effect of its continuously changing pupil on the flatfield during observations. 

Since no flatfield correction was applied, the sky background cannot be estimated naivly by measuring at various locations in the frame away from the galaxy, or by fitting a tilted plane on the basis of small areas far from the galaxy images. In order to determine the sky background we took advantage of its statistical nature; we first created a count-frequency histogram for the galaxy surroundings (that is, in a 3' radius from the galaxy center) and then fitted a Gaussian distribution curve to the peak of the histogram, where the most frequent count value was assumed to represent the sky background. This value was subtracted from the galaxy frame. Due to the nature of the unflatfielded frames we avoided combining frames from different observing nights. Furthermore, only observations with similar seeing values within the same observing night were selected to produce the combined image after  convolution with a Gaussian to match the image seeing.

\section{Results}
\label{txt:results}

As mentioned above, the observations reported here were obtained
during the PV phase of SALT in a ``shared-risk'' mode. This implied
that the telescope was not optimized for imaging, the IQ problem that 
troubled the telescope at that stage of commissioning was not yet solved and
the mirror segments were not freshly aluminized. The implications are that the images outside a $\sim3'$ circle had reduced resolution, and sometimes the seeing
size degraded during the observation yielding some unusable
images.

\begin{figure*}
 \begin{minipage}{130mm}
\subfigure{\includegraphics[width=4cm]{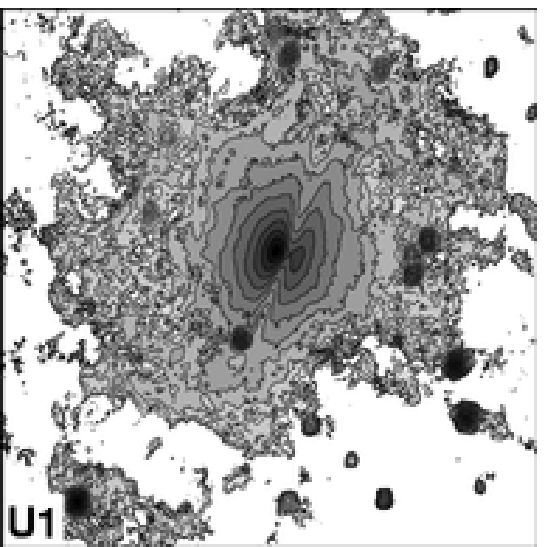}}\vspace{-6.2mm}
\subfigure{\includegraphics[width=4cm]{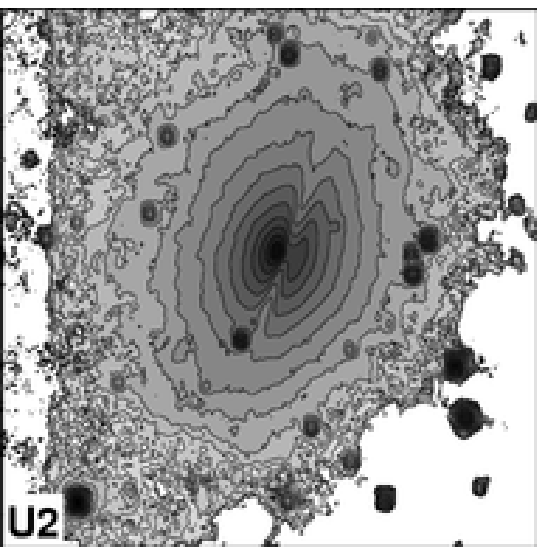}}
\subfigure{\includegraphics[width=4cm]{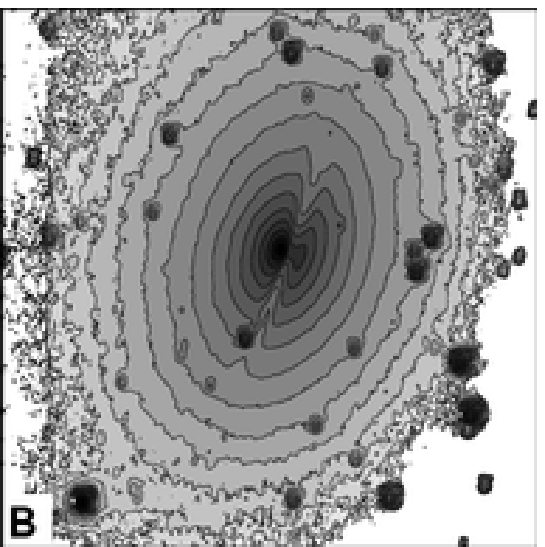}}\newline
\subfigure{\includegraphics[width=4cm]{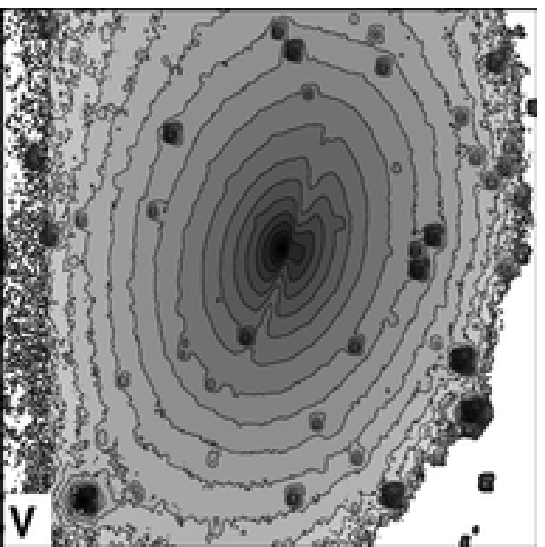}} \vspace{-6.2mm}
\subfigure{\includegraphics[width=4cm]{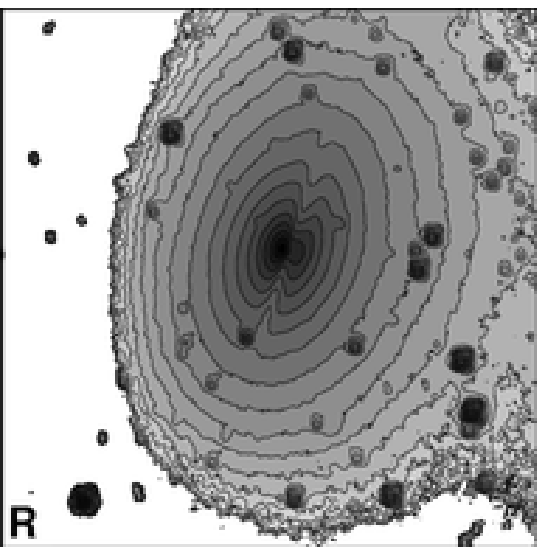}}
\subfigure{\includegraphics[width=4cm]{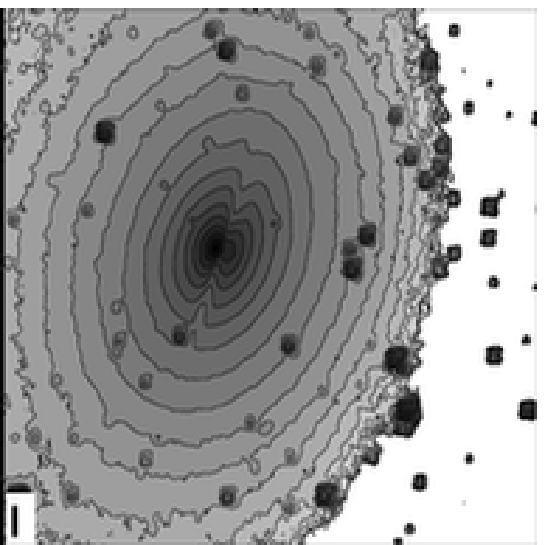}}
\vspace{20pt}
\caption{Filled contour maps for NGC3283 with 0.5 magnitude steps. The images are 112"$\times$112" with north up and east to the left. The dark lane crosses the galactic center almost vertically.}
\label{fig:H1029}
\end{minipage}
\end{figure*}

\begin{figure*}
\begin{minipage}{130mm}
\subfigure{\includegraphics[width=4cm]{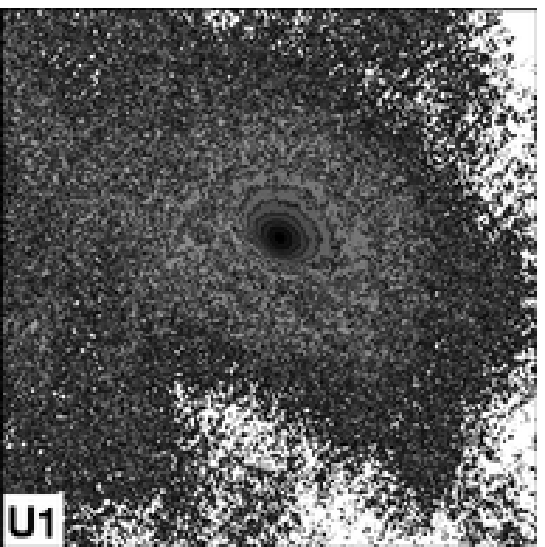}}\vspace{-6.2mm}
\subfigure{\includegraphics[width=4cm]{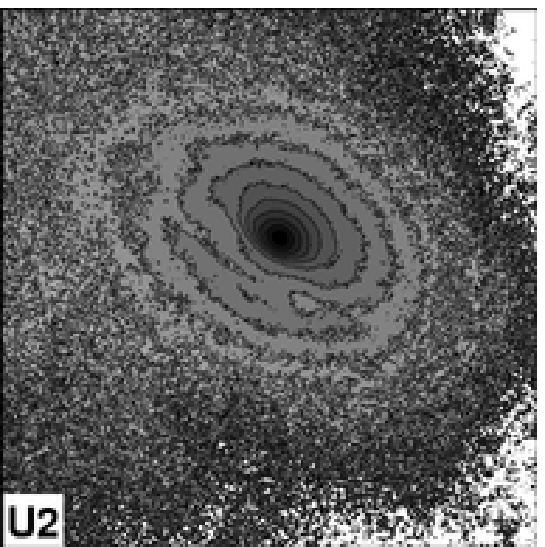}}
\subfigure{\includegraphics[width=4cm]{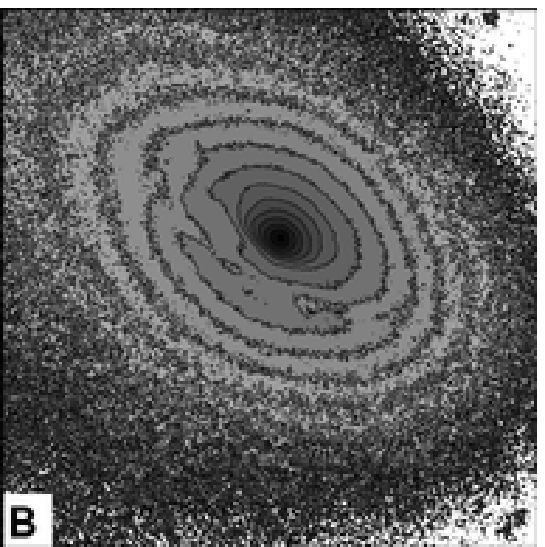}}\newline
\subfigure{\includegraphics[width=4cm]{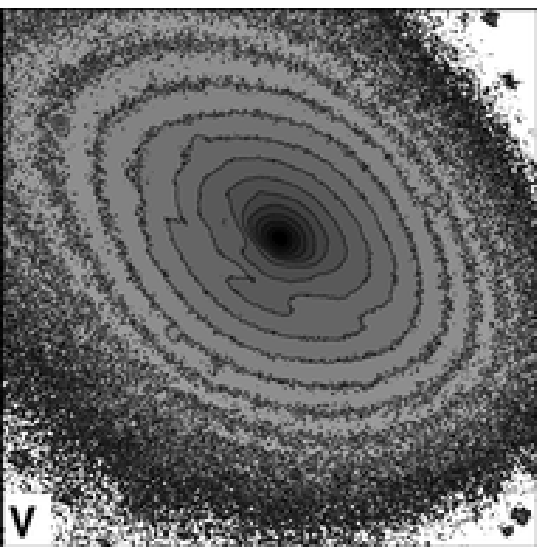}}\vspace{-6.2mm}
\subfigure{\includegraphics[width=4cm]{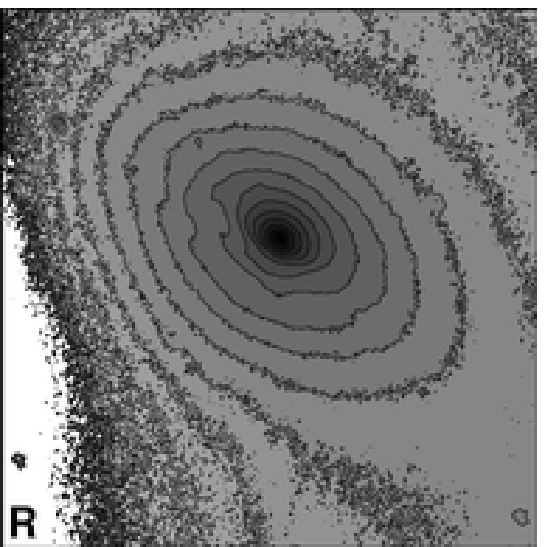}}
\subfigure{\includegraphics[width=4cm]{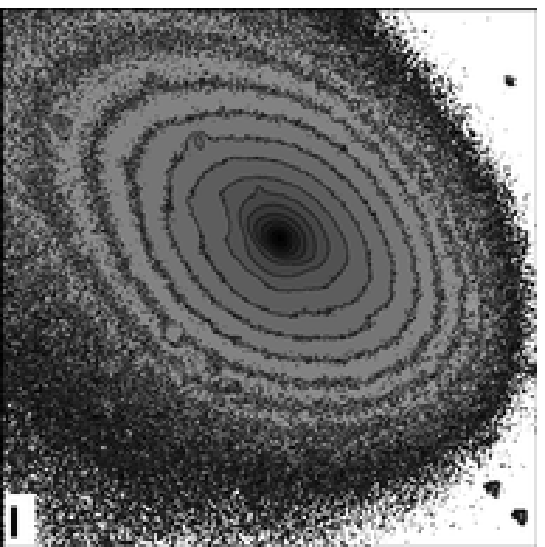}}
\vspace{20pt}
\caption{Filled contour maps for NGC3497 with 0.5 magnitude steps. The images are 112"$\times$112" where north is up and east is left. The dark lane is below and to the left of the galactic center.}
\label{fig:ngc3497}
\end{minipage}
\end{figure*}

\begin{figure*}
 \begin{minipage}{130mm}
\subfigure{\includegraphics[width=4cm]{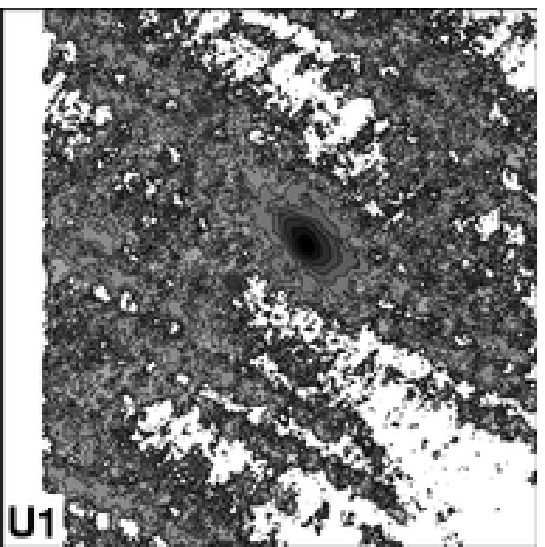}}\vspace{-6.2mm}
\subfigure{\includegraphics[width=4cm]{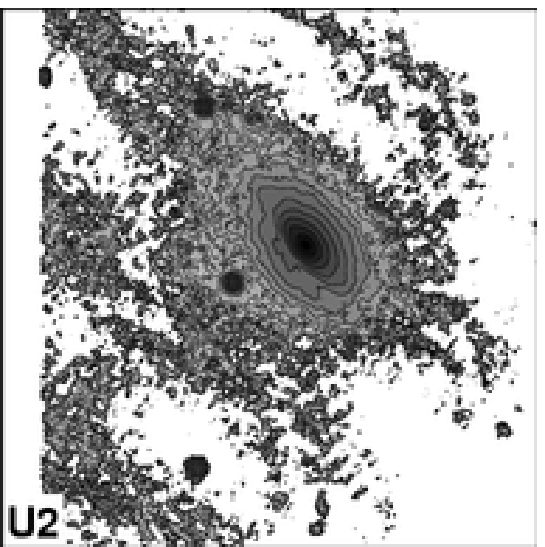}}
\subfigure{\includegraphics[width=4cm]{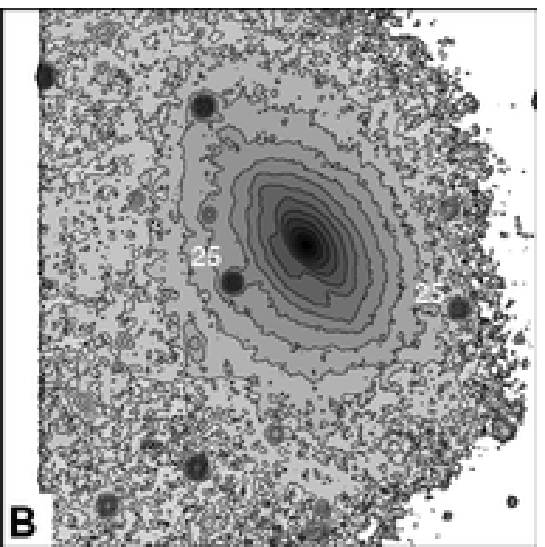}}\newline
\subfigure{\includegraphics[width=4cm]{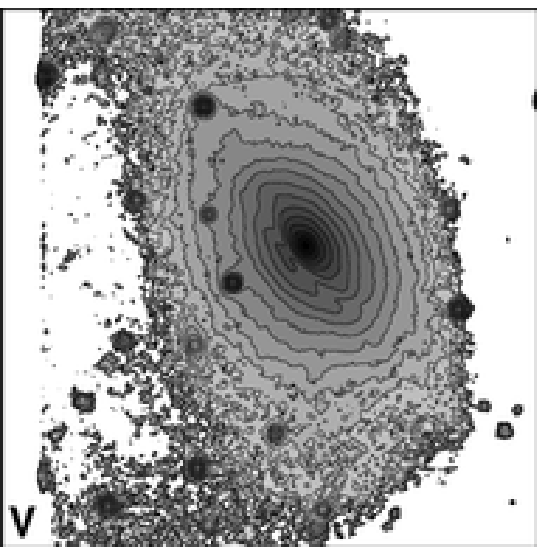}}\vspace{-6.2mm}
\subfigure{\includegraphics[width=4cm]{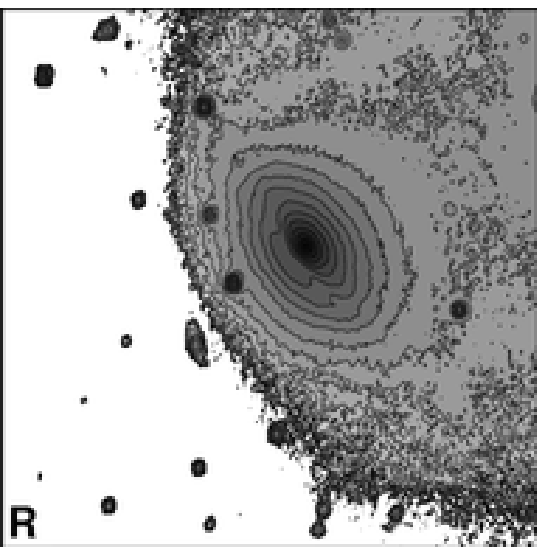}}
\subfigure{\includegraphics[width=4cm]{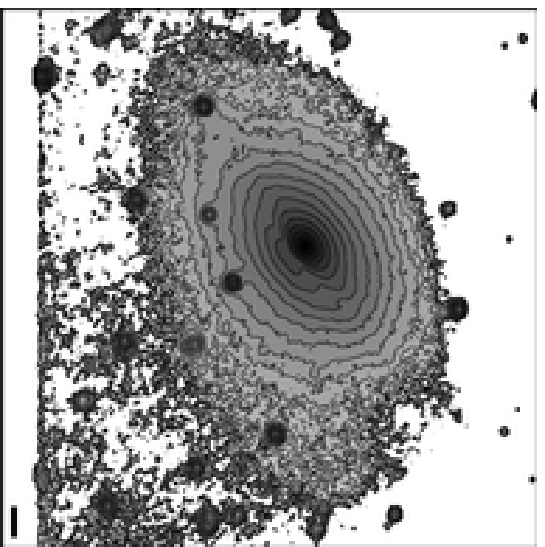}}
\vspace{30pt}
\caption{Filled contour maps for AM1118-290 with 0.5 magnitude steps. The images are 112"$\times$112" where north is up and east is left. The dark lane is below and to the left of the galactic center. The surface magnitude level at 25 mag/arcsec$^2$ is labelled in the B-band image using values from the RC3 catalog.}
\label{fig:H1118}
\end{minipage}
\end{figure*}

\begin{figure*}
 \begin{minipage}{130mm}
\subfigure{\includegraphics[width=4cm]{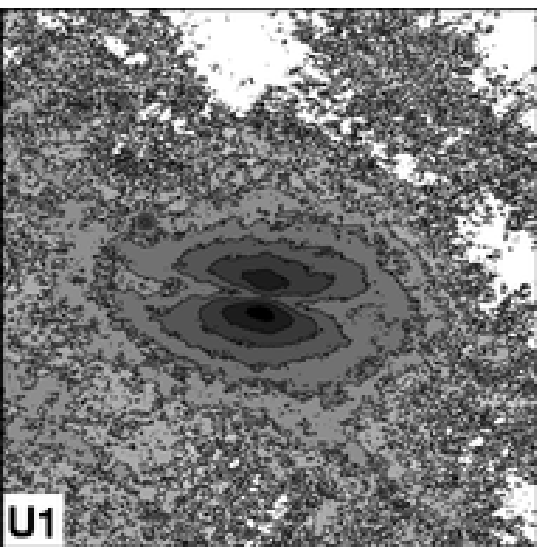}}\vspace{-6.2mm}
\subfigure{\includegraphics[width=4cm]{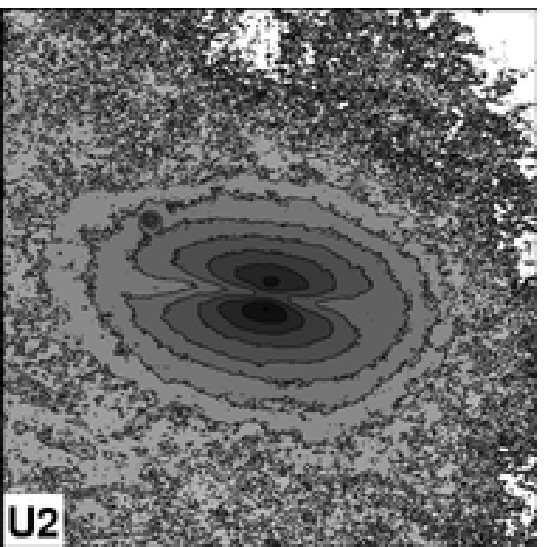}}
\subfigure{\includegraphics[width=4cm]{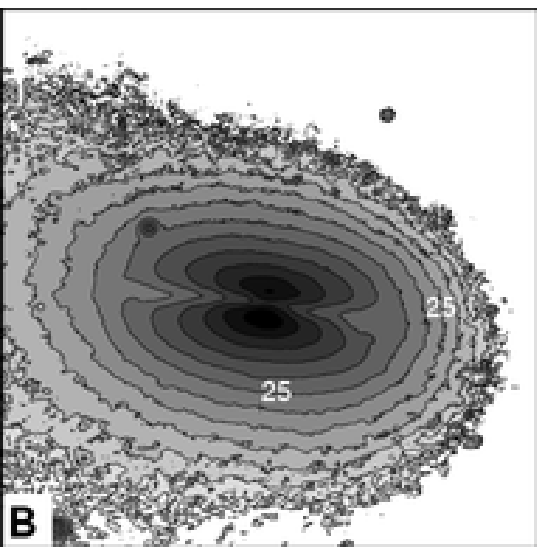}}\newline
\subfigure{\includegraphics[width=4cm]{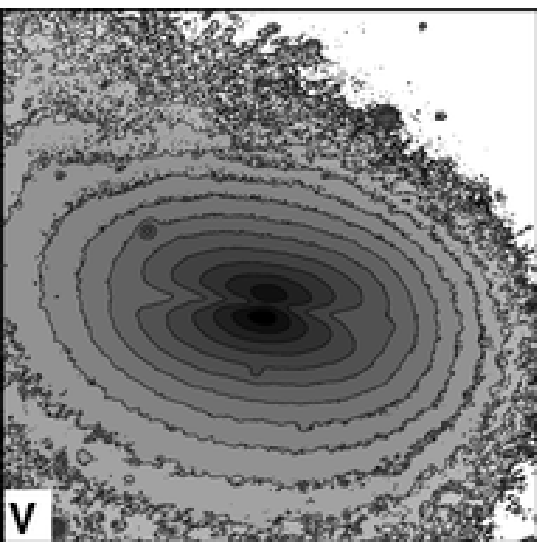}}\vspace{-6.2mm}
\subfigure{\includegraphics[width=4cm]{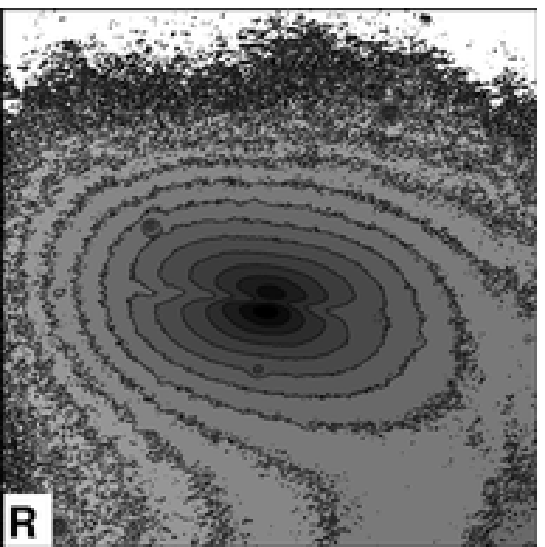}}
\subfigure{\includegraphics[width=4cm]{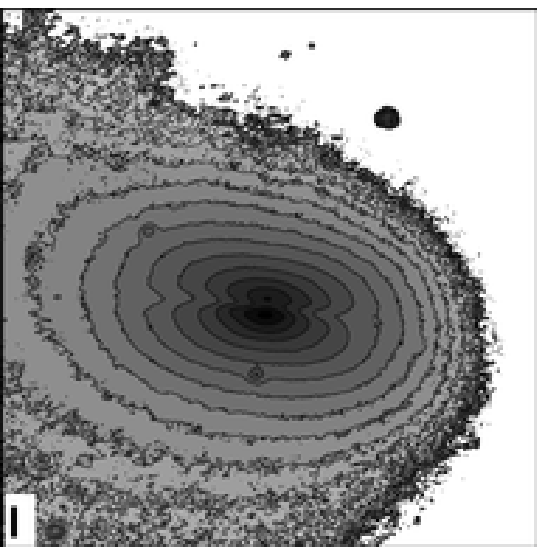}}
\vspace{30pt}
\caption{Filled contour maps for NGC4370 with 0.5 magnitude steps. The images are 112"$\times$112" where north is up and east is left. The dark lane crosses the galactic center almost horizontally. The surface magnitude level at 25 mag/arcsec$^2$ is labelled in the B-band image using values from the RC3 catalog.}
\label{fig:ngc4370}
\end{minipage}
\end{figure*}

\begin{figure*}
 \begin{minipage}{130mm}
\subfigure{\includegraphics[width=4cm]{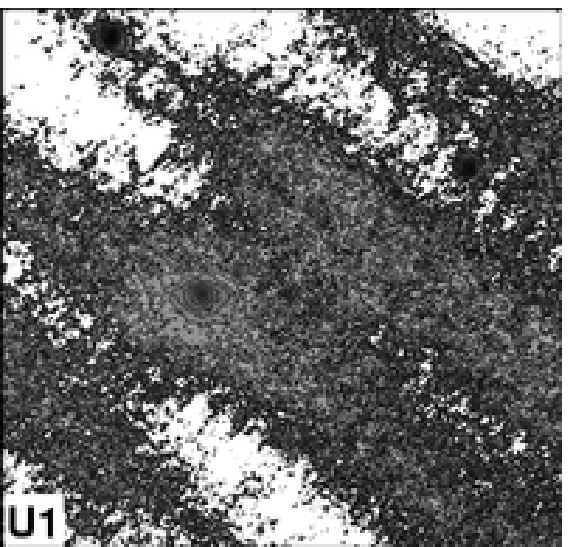}}\vspace{-6.2mm}
\subfigure{\includegraphics[width=4cm]{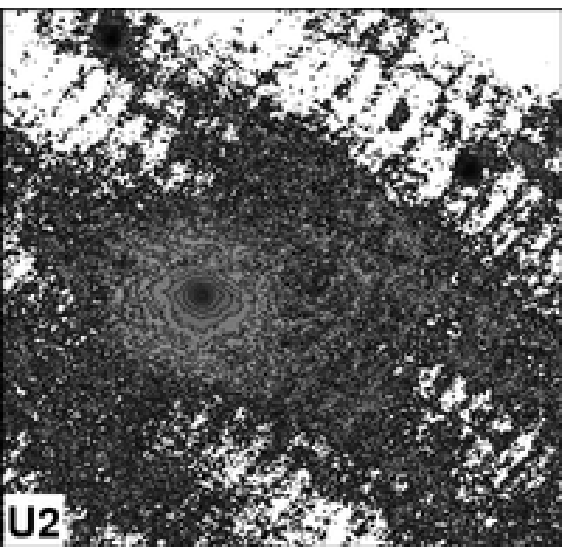}}
\subfigure{\includegraphics[width=4cm]{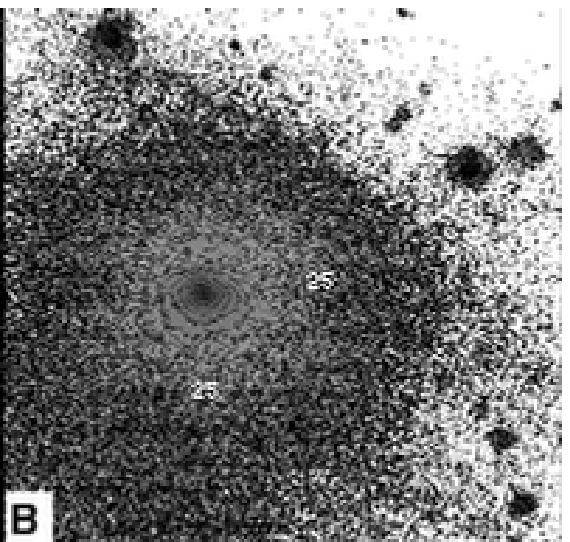}}\newline
\subfigure{\includegraphics[width=4cm]{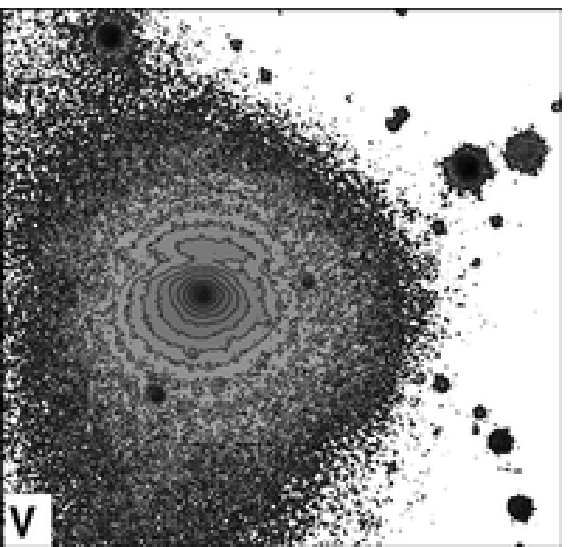}}\vspace{-6.2mm}
\subfigure{\includegraphics[width=4cm]{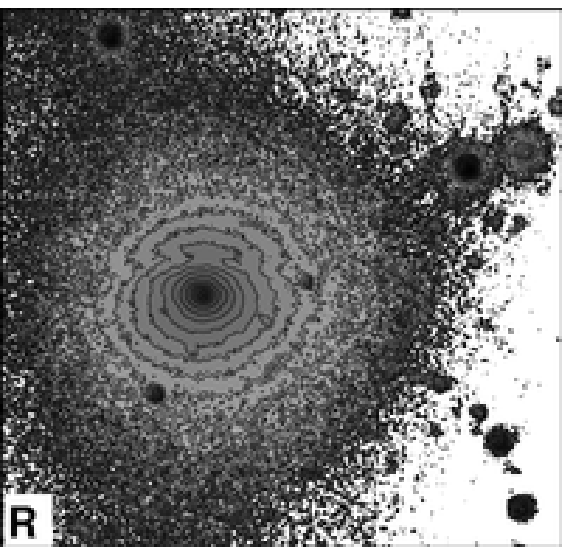}}
\subfigure{\includegraphics[width=4cm]{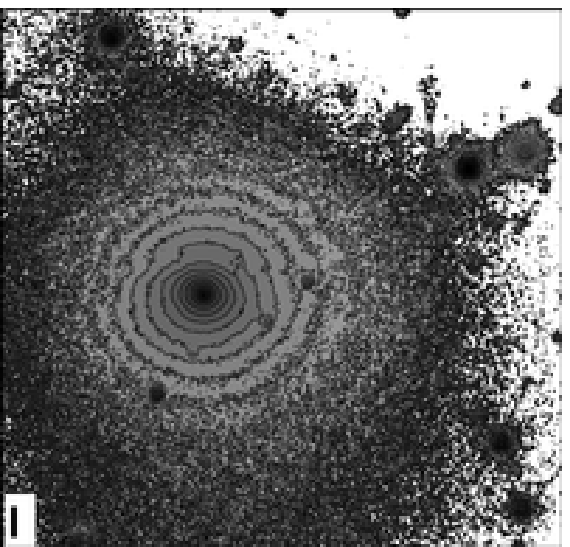}}
\vspace{30pt}
\caption{Filled contour maps for AM1352-333 with 0.5 magnitude steps. The images are 112"$\times$112" where 
north is up and east is left. The dark lane crosses above the galactic center horizontally.}
\label{fig:H1352}
\end{minipage}
\end{figure*}

\begin{figure*}
 \begin{minipage}{130mm}
\subfigure{\includegraphics[width=4cm]{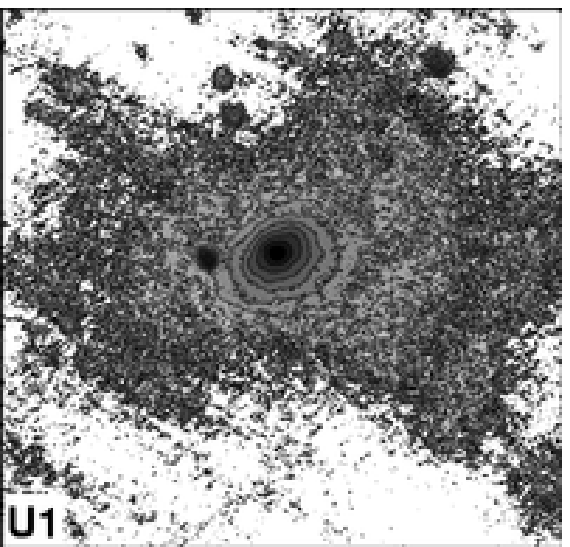}}\vspace{-6.2mm}
\subfigure{\includegraphics[width=4cm]{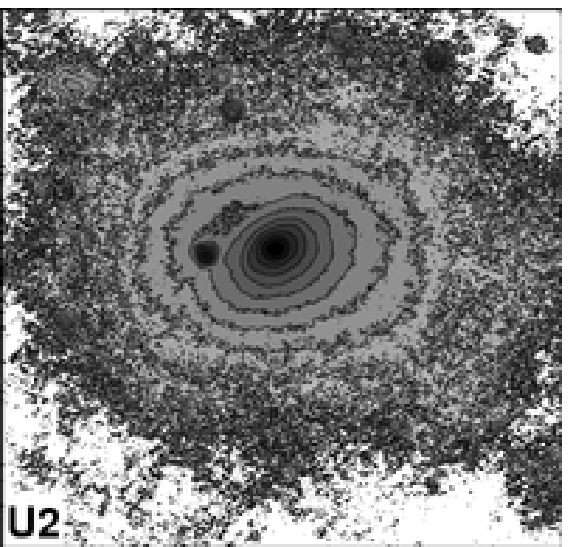}}
\subfigure{\includegraphics[width=4cm]{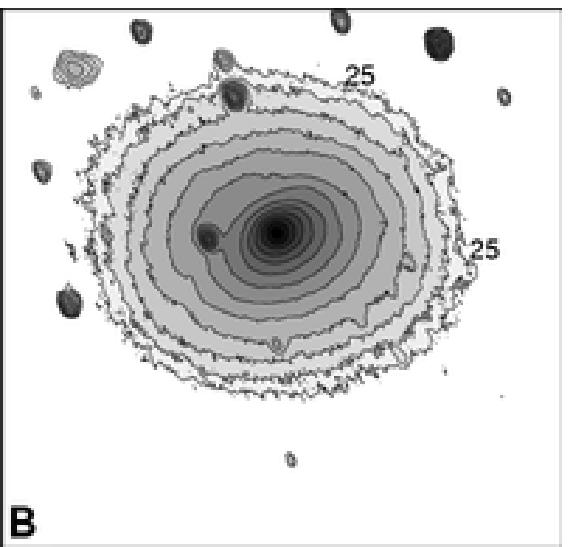}}\newline
\subfigure{\includegraphics[width=4cm]{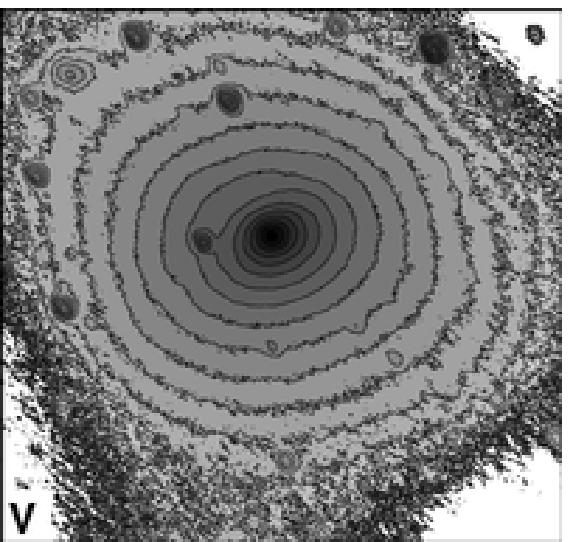}}\vspace{-6.2mm}
\subfigure{\includegraphics[width=4cm]{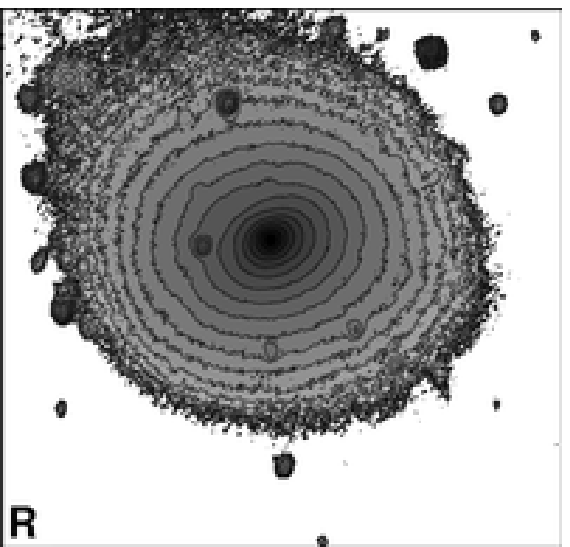}}
\subfigure{\includegraphics[width=4cm]{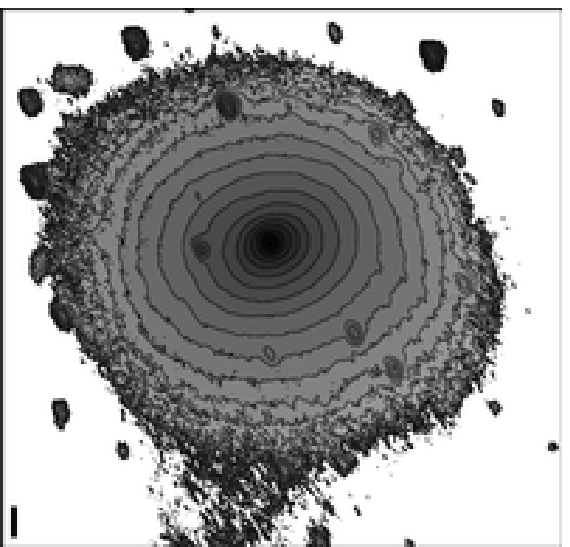}}
\vspace{30pt}
\caption{Filled contour maps for NGC5626 with 0.5 magnitude steps. The images are 112"$\times$112" where north is up and east is left. The dark lane is above and to the left of the galactic center. The surface magnitude level at 25 mag/arcsec$^2$ is labelled in the B-band image using values from the RC3 catalog.}
\label{fig:ngc5626}
\end{minipage}
\end{figure*}

\begin{figure*}
 \begin{minipage}{130mm}
\subfigure{\includegraphics[width=4cm]{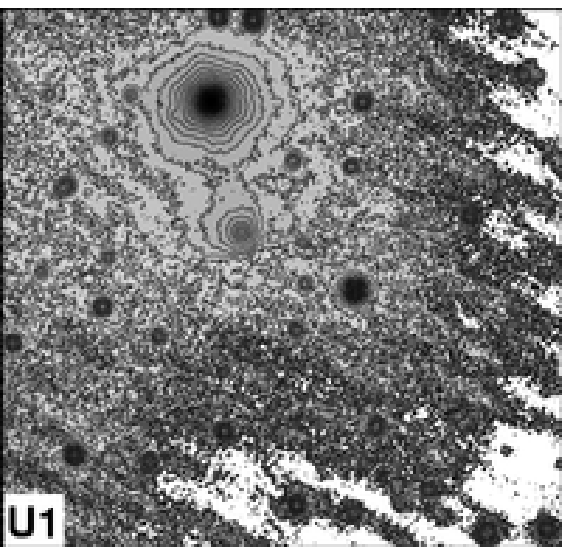}}\vspace{-6.2mm}
\subfigure{\includegraphics[width=4cm]{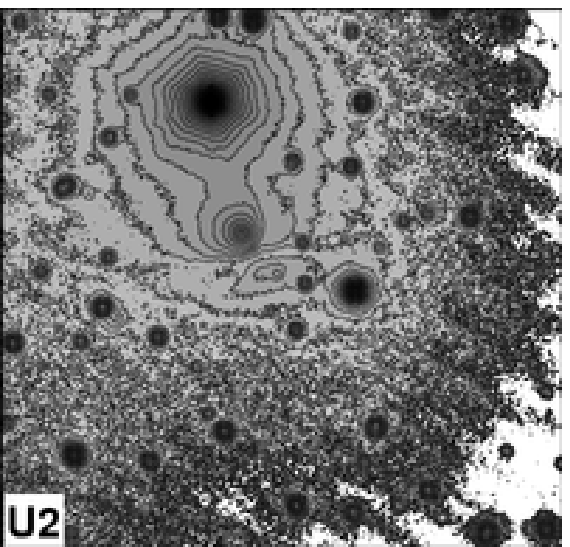}}
\subfigure{\includegraphics[width=4cm]{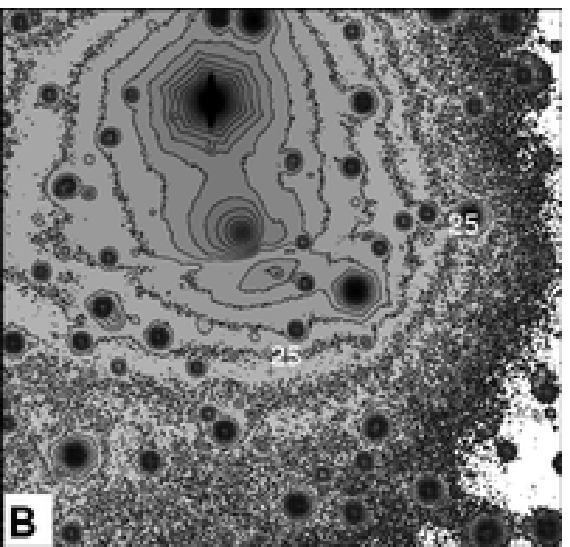}}\newline
\subfigure{\includegraphics[width=4cm]{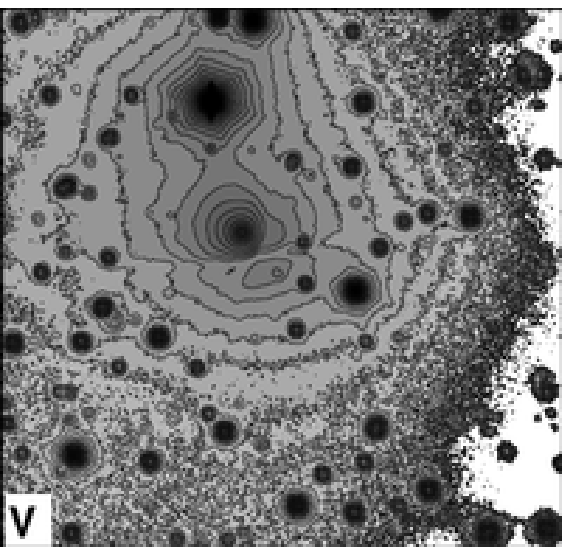}}\vspace{-6.2mm}
\subfigure{\includegraphics[width=4cm]{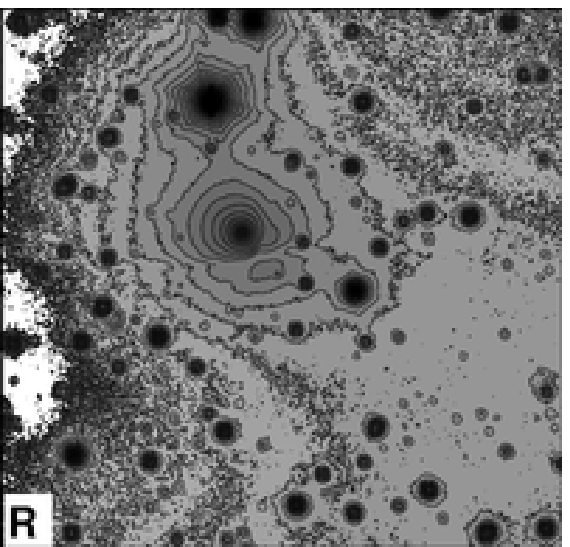}}
\subfigure{\includegraphics[width=4cm]{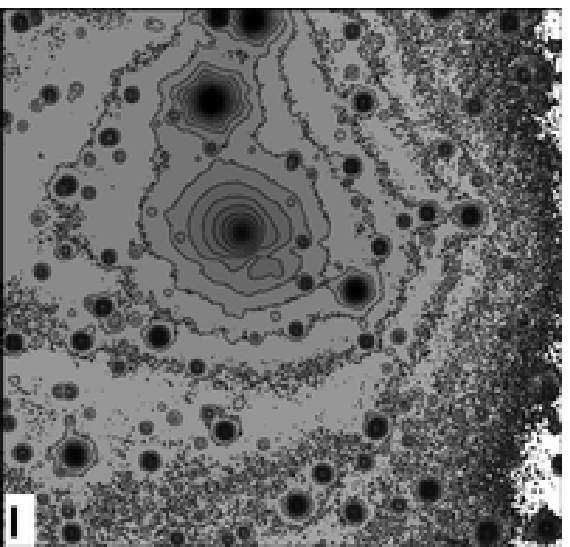}}
\vspace{30pt}
\caption{Filled contour maps for AM1459-722 with 0.5 magnitude steps. The images are 112"$\times$112" where north is up and east is left. The galaxy is centered in these images and the dark lane crosses horizontally below the galactic center from. The bright object above the galaxy is a nearby bright star. The surface magnitude level at 25 mag/arcsec$^2$ is labelled in the B-band image using values from the RC3 catalog.}
\label{fig:H1459}
\end{minipage}
\end{figure*}

\begin{figure*}
 \begin{minipage}{130mm}
\subfigure{\includegraphics[width=4cm]{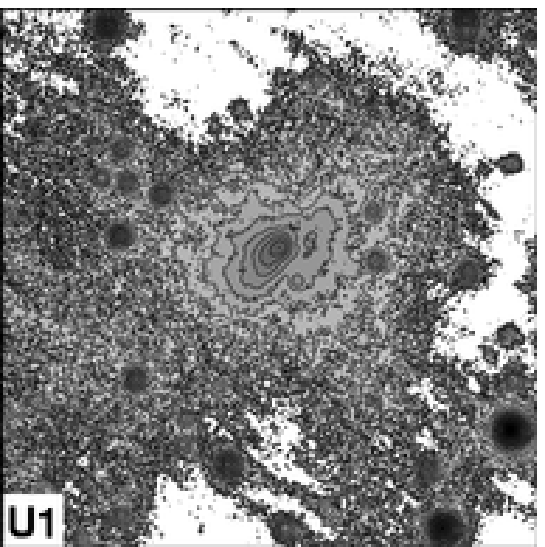}}\vspace{-6.2mm}
\subfigure{\includegraphics[width=4cm]{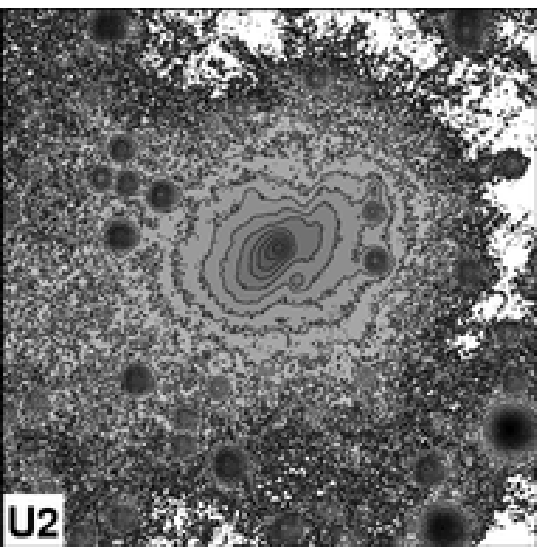}}
\subfigure{\includegraphics[width=4cm]{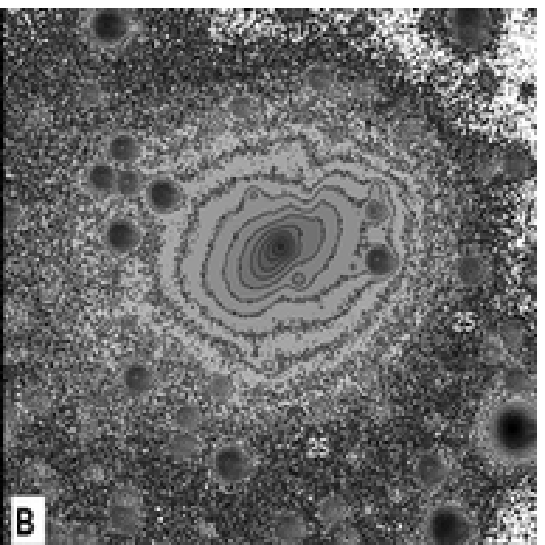}}\newline
\subfigure{\includegraphics[width=4cm]{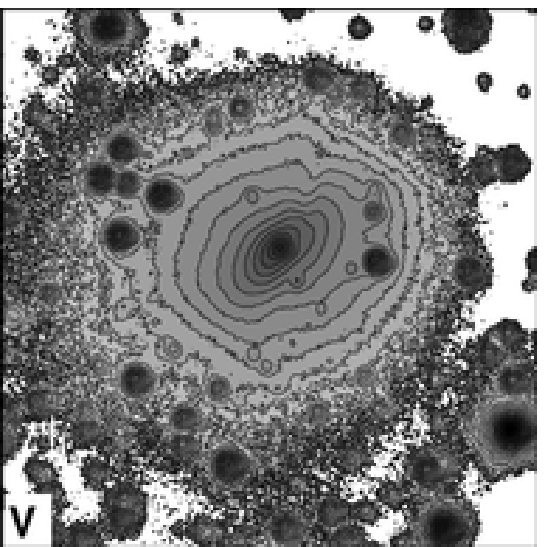}}\vspace{-6.2mm}
\subfigure{\includegraphics[width=4cm]{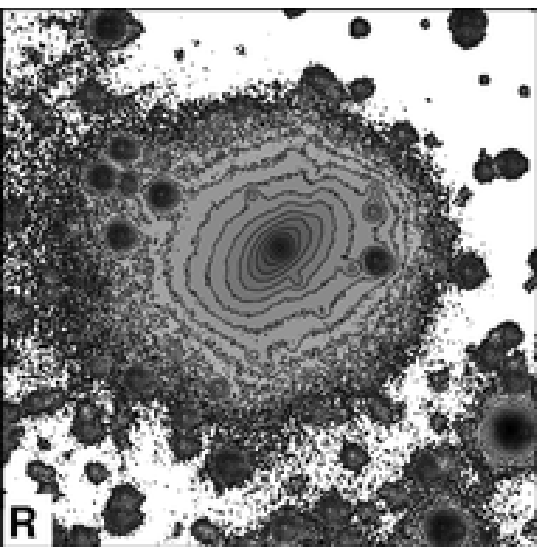}}
\subfigure{\includegraphics[width=4cm]{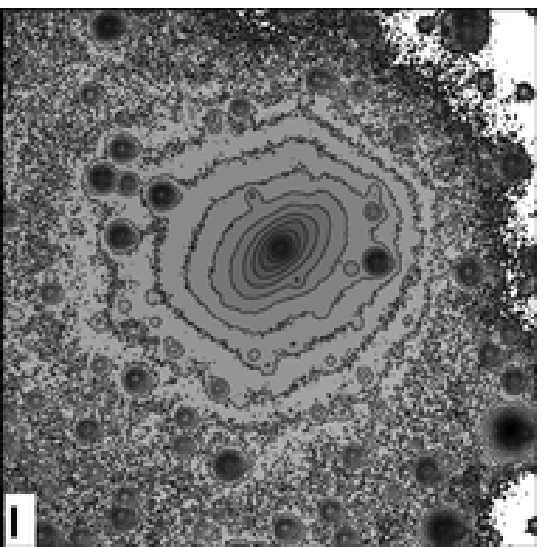}}
\vspace{30pt}
\caption{Filled contour maps for NGC5799 with 0.5 magnitude steps. The images are 112"$\times$112" where north is up and east is left. One dark lane is below the galactic center while another one is to the right of the galactic center. The surface magnitude level at 25 mag/arcsec$^2$ is labelled in the B-band image using values from the RC3 catalog.}
\label{fig:ngc5799}
\end{minipage}
\end{figure*}

\begin{figure*}
 \begin{minipage}{130mm}
\subfigure{\includegraphics[width=4cm]{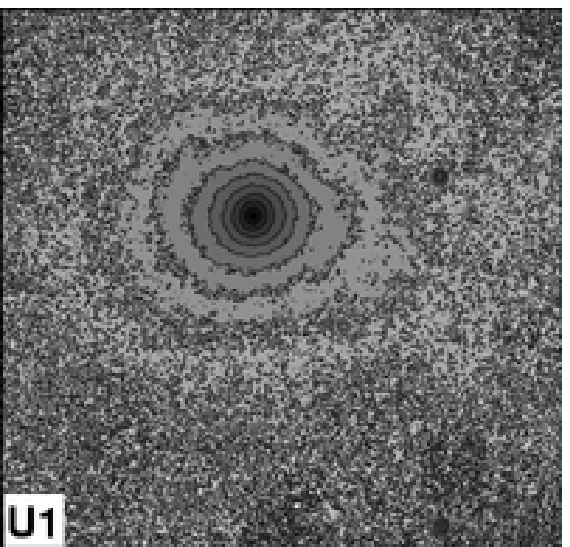}}\vspace{-6.2mm}
\subfigure{\includegraphics[width=4cm]{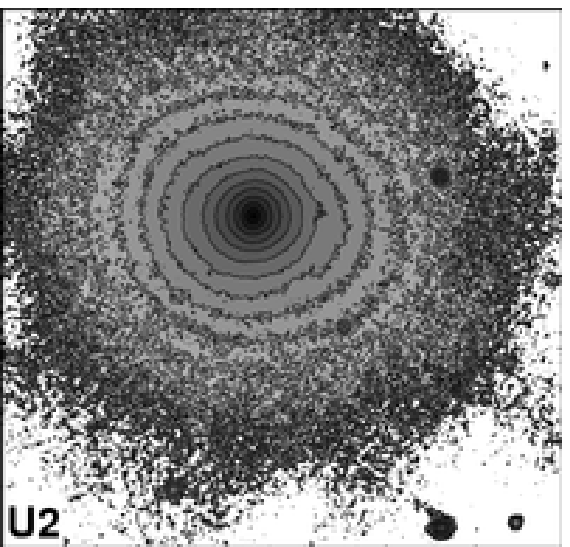}}
\subfigure{\includegraphics[width=4cm]{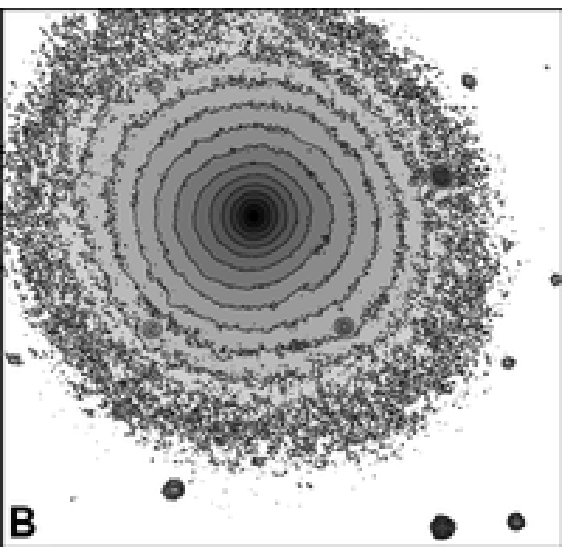}}\newline
\subfigure{\includegraphics[width=4cm]{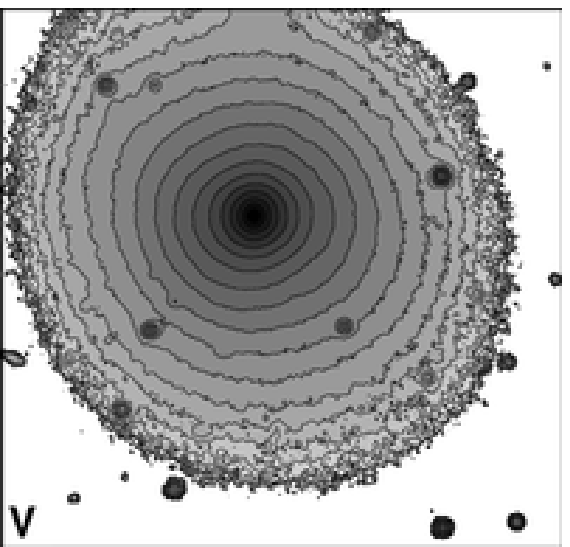}}\vspace{-6.2mm}
\subfigure{\includegraphics[width=4cm]{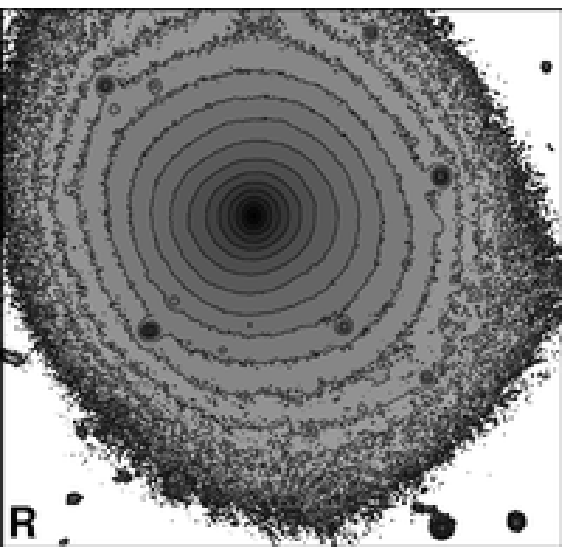}}
\subfigure{\includegraphics[width=4cm]{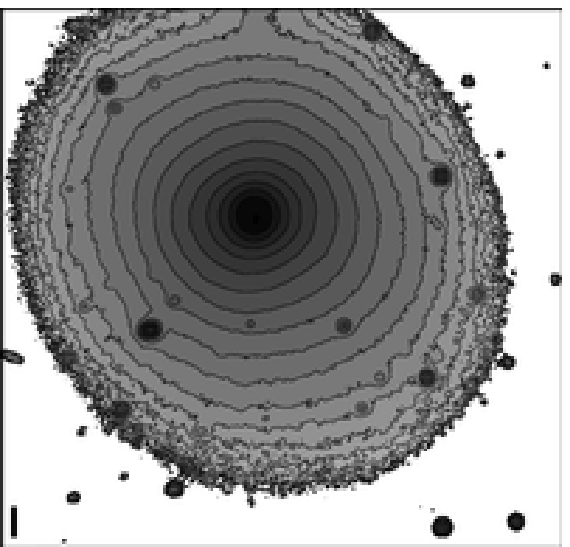}}
\vspace{30pt}
\caption{Filled contour maps for NGC5898 with 0.5 magnitude steps. The images are 112"$\times$112" where north is up and east is left. The dark lane is to the right of the galactic center.}
\label{fig:ngc5898}
\end{minipage}
\end{figure*}

We present contour maps corresponding to the various filters in use for each galaxy.
Figs.\ \ref{fig:H1029} to \ref{fig:ngc5898} show filled contour maps for each of the sample galaxies. The contour maps are created from the reduced and combined final images and are each plotted with 0.5 magnitude steps. 
As evident from the contour maps, the U1 and U2 images are significally fainter than the optical band images, producing at the dark lanes a S/N ratio of $\sim 10$ and $\sim 25$ per resolution element, respectively. This, while the characteristic S/N ratio per resolution element for the B, V, R \& I images may reach values of $\sim 100$ and even higher. Since the images are not calibrated, the estimation of the contour level where the surface magnitude reaches to 25 mag/arcsec$^2$ is based on the semi-major axes measurements from the RC3 catalog (see de Vaucouleurs et al.\ 1992). The corresponding B-image contours are labelled, allowing for a comparison between the S/N ratio in different filters for each galaxy. Galaxies where the major-axis is too large to fit the 112"x112" contour maps are not labelled.
Some of the contour maps, mainly for the near-UV bands, show low S/N chevron vintage patterns due to both electronics and the unflatfielded nature of the images. In some cases, stellar images in Figs.\ \ref{fig:H1029} to \ref{fig:ngc5898} appear somewhat distorted mainly towards the edges, probably caused by the IQ problems mentioned earlier.
We also present B-I colour-index maps of the sample galaxies in Fig.\ \ref{fig:BImaps}. These images show the morphological differences between the suspected dust lanes, presenting concentric rings of dust (e.g., NGC3497 and NGC5626), warped dust disks (e.g., NGC4370) and an orthogonal set of dust lanes (AM1459-722). 


\section{Analysis}
\label{txt:analyse}
In order to estimate the dust extinction in galaxies one can compare the light distribution in the observed galaxy with its dust free model.
Since the target galaxies are early-type, one could assume that they are
spheroids and use a de Vaucouleurs (1948) light distribution
\begin{equation}
\mu\left(r\right)=\mu_e + 8.325\left[\left(\frac{r}{r_e}\right)^{1/4}-1\right] 
\label{eq:devaucouleurs},
\end{equation}
or a more flexible fit using a S\'{e}rsic light distribution for a
bulge which would fit an elliptical galaxy
\begin{equation}
\mu_b\left(x,y\right)=\mu_e+k_n\left[\left(\frac{r_b}{r_e}\right)^{1/n}-1\right] 
\label{eq:disk},
\end{equation}
where $r_b$=$\left[ |x|^{c+2}+|\frac{y}{q_b}|^{c+2}\right] ^{\frac{1}{\left( c+2\right)}}$, $q_b$ is the axial ratio of the bulge and a pure ellipse has c=0.
A more realistic fit assumes that the galaxy is a
superposition of a bulge following the S\'{e}rsic law and an
exponential disk, whose light distribution is given by
\begin{equation}
\mu_d\left(x, y\right)=\mu_0+1.086\left(\frac{r_d}{h}\right),
\end{equation}
where $r_d$=$\left[ |x|^{c+2}+|\frac{y}{q_d}|^{c+2}\right] ^{\frac{1}{\left( c+2\right)}}$, $q_d$ is the axial ratio of the disk and a pure ellipse has c=0. Such an assumption would fit a lenticular galaxy.

One method of fitting a smooth galaxy model to the images is to use the ``standard'' model mentioned above for the entire galaxy image, which may be preceded by the masking of parts of the image where obvious dust is present.
It has also been shown that constructing a dust-free model for an early-type galaxy, which has a fairly smooth and symmetric light distribution with respect to its nucleus, can be easily done by fitting ellipses to the isophotes of the observed image. Although this method has been used succesfully in a number of cases (Brosch and Loinger 1991; Goudfrooij et al.\ 1994; Sahu et al.\ 1998 and Patil et al.\ 2007), it seems to be less effective in our case. First, in most cases, the masking of the entire dusty region in the elliptical fits leads to a percentage of masked pixels that is too large to be handled by the ellipse-fitting routine in IRAF. Second, it seems that fitting ellipses for the images in the near-UV produces poor results mainly due to the characteristic relatively low S/N ratio per resolution element (which can even be as low as 5) following the background subtraction. For these reasons we selected a different fitting method that is described below.

\subsection{Extinction maps} \label{S:extinctionmaps}
The appearance of the dust lane, regardless of its orientation with respect to the stellar body, can be described in several ways (Bertola 1987). The dust lane may lie exactly along one of the axes of the stellar body, presumably representing a disk seen edge-on, or show a slight curvature, suggesting a disk seen almost edge-on. Dust lanes may also appear as full rings or may appear to be warped at their outer parts where the minor-axis dust lane bends toward the major axis in the outer regions. In a few cases, a set of multiple parallel dust lanes is present, suggesting a system of coplanar rings seen at an angle. Considering the dust morphology, one should also consider the extent of the dust lane with respect to the galaxy, i.e., whether it is confined to the innermost regions of the galaxy or whether it follows the luminous stellar material down to its detection limit.

While the  dust lanes do not always appear to be symmetric under reflection through the nucleus, we may assume that the light distribution of the underlying galaxy does hold such symmetry properties in order to extract the extinction maps. The extinction at each location within a resolution element may be measured using
\begin{equation}
A_{\lambda}=-2.5 \, log\left[\frac{I_{\lambda,obs}}{I_{\lambda,mod}}\right]
\end{equation}
which under reflection symmetry may be expressed as
\begin{equation}
A_{\lambda}=-2.5\sum\limits_{ij}\left[logI_{\lambda}\left(x_i,y_j\right)-logI_{\lambda}\left(-x_i,-y_j\right)\right]
\end{equation}
where $A_{\lambda}$ represents the amount of extinction in a particular band on a magnitude scale, and $I_{\lambda}\left(x,y\right)$ represents the average intensity of a rectangular seeing-size box at location $(x,y)$ away from the nucleus. Such rectangular boxes were translated over the dust-occupied regions of each galaxy with no overlap, though the nuclear regions (radius$\leq$5{"}) were excluded to avoid seeing-related effects. We also assume the Galactic extinction in the line of sight is uniform over each galaxy, thus no correction for Galactic extinction is necessary. 

However, the reflection symmetry does not necessarily hold when major-axis dust-lane galaxies are concerned. 
Three of our sample galaxies (NGC3283, NGC3497 and NGC4370) were studied by Bertola et al.\ (1988) as part of their study of major-axis dust-lane ellipticals. While from the B-I colour-index maps it appears that the dust lane in NGC3497 does not lie exactly along the major axis (see also Patil et al.\ 2007), we do identify a major-axis dust lane in NGC4370 and NGC3283. We should note here that there is a tendency in the usual morphology classification schemes to call S0 any galaxy where a stellar body is crossed by a dust lane, regardless of its orientation.

In order to extract an extinction map for each of these two galaxies we used a two-dimensional bulge/disk decomposition analysis code (BUDDA; Souza, Gadotti \& dos Anjos 2004) to derive the galaxies' global parameters and to build 2D models for the galaxies. In order to determine initial values for the galactic center, position angle, ellipticity and ellipse index (i.e., the galaxy boxiness), we fitted elliptical isophotes to the outer regions of the galaxies, which are least affected by dust. Initial values for the effective radius and effective intensity (see eq. \ref{eq:devaucouleurs}) were obtained by fitting a de Vaucouleurs profile to the minor axis, which was first rotated to lie parallel to one of the x-y axes for convenience. Excluding the inner regions occupied by dust, this profile proved to fit well the light distribution along the minor axis, supporting the assumption that these galaxies are indeed ellipticals. 
This fit was initialy performed for the optical band images in an iterative mode, meaning that dust regions were excluded from each fit according to the residual image based on the previous fit, until revealing what seems to be the full extent of the dust lanes.
For consistency, we compared the global parameters obtained for each band, including the location of the galactic center, the position angle, the ellipticity, the ellipse index and also the effective radius. These values were found to match and were later used to fit models for the U1 and U2 images, where only the effective intensity was left as a free parameter due to the relatively low S/N ratio of these images.

\subsection{Extinction curves}
\label{S:extinctionval}
As explained in \S~\ref{S:extinctionmaps}, total extinction values were measured for different dust-extinguished regions in each galaxy. Those values were later used to derive the extinction values for each galaxy by fitting a linear regression between the total extinction $A_{\lambda}$ and the selective extinction E(B-V)=$A_B-A_V$ using weighted least-squares with a fixed zero extinction point. The best-fitting slopes were used as the extinction values $R_{\lambda}\left( {\equiv}\frac{A_\lambda}{E\left( B-V\right)} \right) $ for each galaxy and were compared with the Galactic values. 
We also used this method to fit a linear regression between the total extinction at wavelength $\lambda$ ($A_\lambda$) and the optical extinction $A_V$. Extinction values normalized to the optical extinction were obtained by averaging the best-fitted slope of $A_{\lambda}$ versus $A_V$ and the reciprocal slope of $A_V$ versus $A_{\lambda}$.
Extinction values obtained using the latter normalization seem to produce less noisy profiles, probably as a result of the larger relative errors of E(B-V) compared to those of $A_V$, where the B image of an early-type galaxy is usually significally fainter than the V image. 

The derivation of the dust extinction law using the procedure described above may be affected by the location of the dust within the galaxy.
While in our Galaxy we assume all the dust is between us and the light source, this may not be the case for dust lanes or dust patches in galaxies where dust is located within the stellar body. This phenomenon can be illustrated using two simple models developed by Walterbos \& Kennicutt (1988) for two idealized cases. In one case, an optically and geometrically thin dust lane is embedded in the galaxy, thus the observed extinction $A_\lambda$ can be expressed as:
\begin{equation}
A_\lambda=-2.5 \, \mbox{log} \left[ p+\left( 1-p\right)e^{-{\tau}_{\lambda}} \right] 
\label{eq:case1}
\end{equation}
where $p$ is the fraction of light emitted at wavelength $\lambda$ that originates between us and the dust, and $\tau_\lambda$ is the optical depth of the dust expressed in units of $\tau_V$, the optical depth in the V-band, using the Galactic reddening law to make the transformation. In the other simplified case, the dust and stars are homogeneously mixed in the dust lane, thus $A_\lambda$ can be expressed as:
\begin{equation}
A_\lambda=-2.5 \, \mbox{log} \left( \dfrac{1-e^{-\tau_\lambda}}{\tau_\lambda}\right) 
\label{eq:case2}
\end{equation}
where $\tau_\lambda$ is computed on the same basis as in eq.\ \ref{eq:case1}.
Considering eq.\ \ref{eq:case1} and eq.\ \ref{eq:case2}, these two effects seem to increase the slope of the observed extinction curve and the observed extinction values with respect to their intrinsic values in the optical part of the spectrum (see, for example, Brosch et al.\ 1990). This deviation from the intrinsic extinction curve is more significant for larger optical depths.

\begin{figure}
\subfigure{\includegraphics[width=7cm]{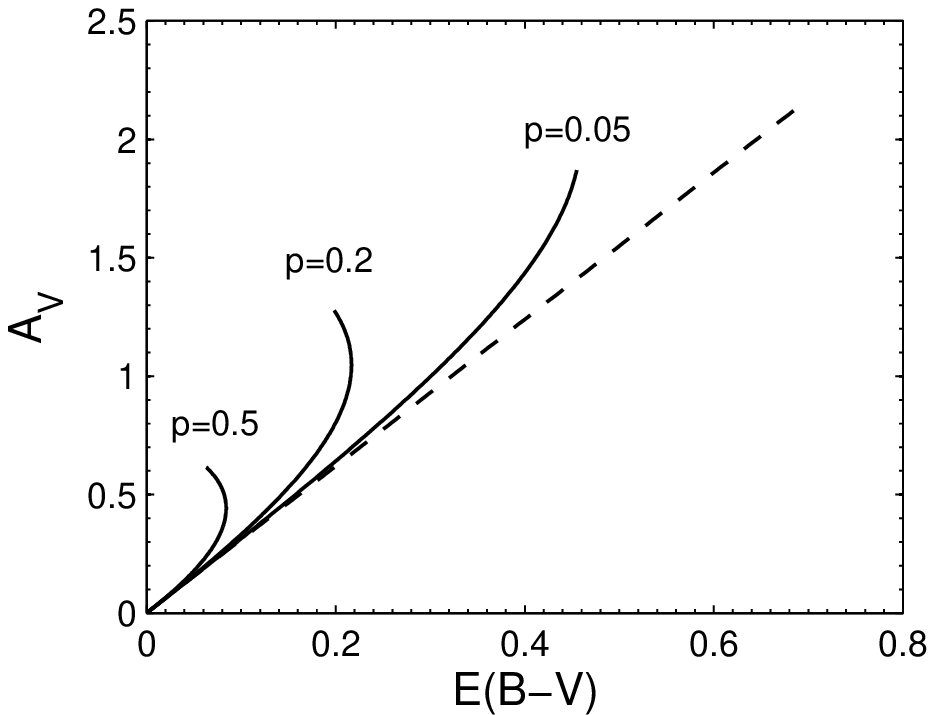}}\vspace{-6.2mm}
\subfigure{\includegraphics[width=7cm]{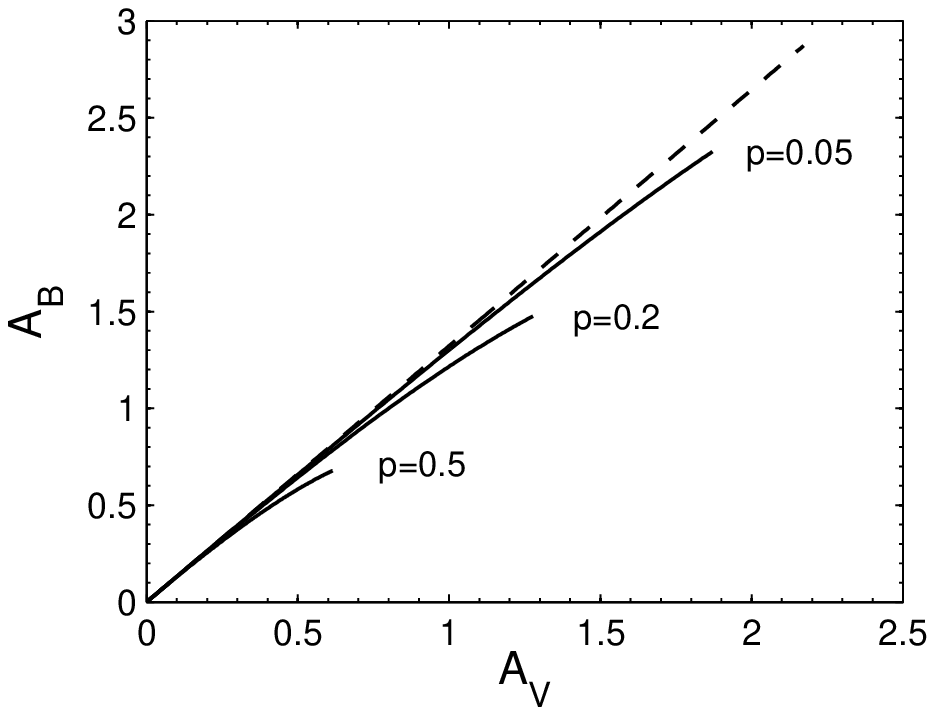}}\newline
\caption{Effects of foreground light on the observed extinction. The dashed line is the typical S\&M relation while the solid lines are for $p=0.05$, $p=0.2$ and $p=0.5$ values. Optical depth values run between $\tau_V=0$ and $\tau_V=2$.}
\label{fig:EvsA}
\end{figure}

\begin{figure}
\subfigure{\includegraphics[width=7cm]{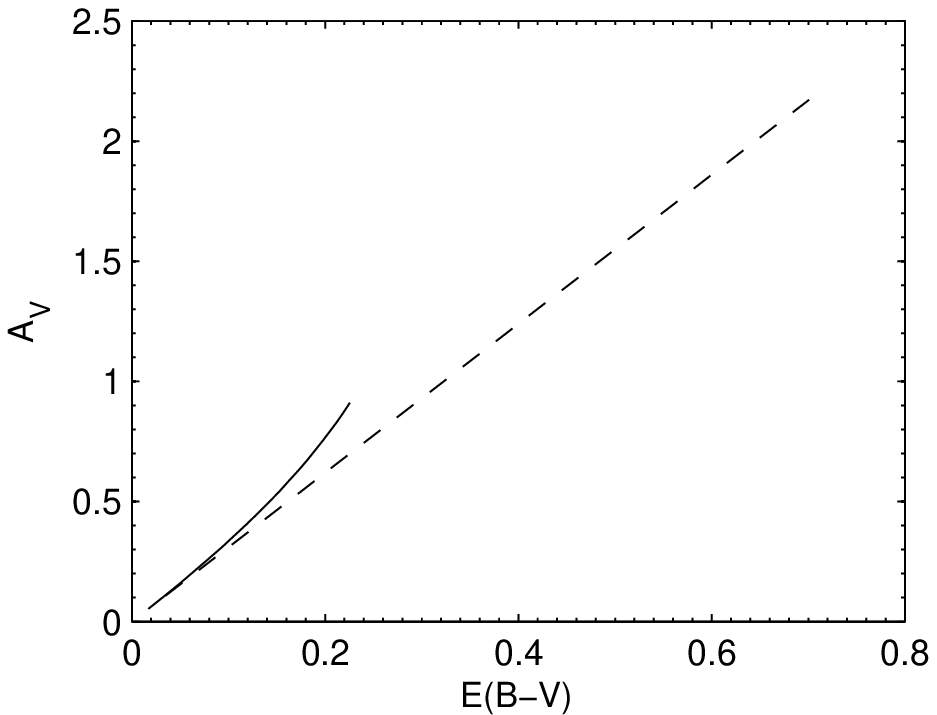}}\vspace{-6.2mm}
\subfigure{\includegraphics[width=7cm]{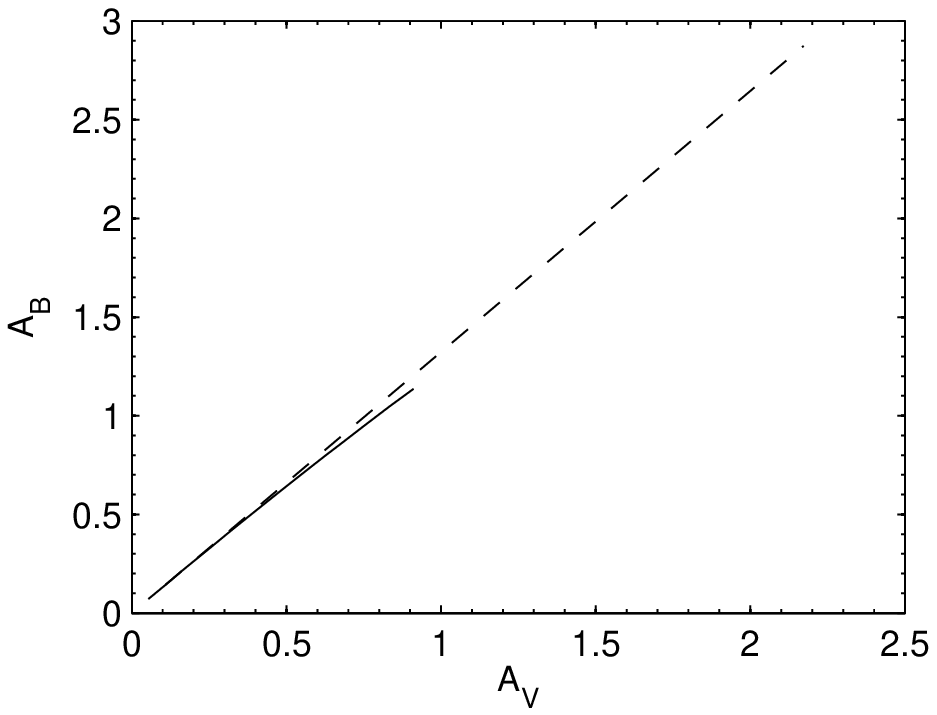}}\newline
\caption{Effect of homogeneously mixed dust and stars on the observed extinction. The dashed line is the typical S\&M relation while the solid line is for the embedded dust. Optical depth values run between $\tau_V=0$ and $\tau_V=2$.}
\label{fig:EvsA2}
\end{figure}

It is also interesting to study the influence of the dust location on the observed relation between $A_\lambda$ and E(B-V) for the case of varying optical depths, as may be found in different regions occupied by dust. The relation $A_\lambda=1.086 \, \tau_\lambda$ is valid as long as we assume the dust is a foreground screen. Since for a given wavelength and a given dust grain distribution the optical depth may only change with varying column density and column length (see also \S~\ref{txt:interp}), this relation implies an expected linear correlation between $A_\lambda$ and E(B-V) where the slope represents the extinction value $R_\lambda$. However, even relatively small amounts of light originating between us and the the dust lane may significantly affect the observed $A_\lambda$ to $\mbox{E(B-V)}$ relation, as illustrated in Fig.\ \ref{fig:EvsA} and Fig.\ \ref{fig:EvsA2}. Therefore, one should be careful when fitting a linear regression to obtain an extinction value, where in the case presented above this linear relation may not be valid even as a first approximation.  In addition to the $A_\lambda$ versus E(B-V) relation, which appears non-linear for relatively small values of $p$ and small optical depths, Fig.\ \ref{fig:EvsA} also shows a similar effect on the $A_\lambda$ versus $A_V$ relation, although as the figure shows, the deviation from linearity becomes significant only for high $p$ values and very high optical depths. 

\begin{figure*}
\begin{minipage}{150mm}
\subfigure{\includegraphics[width=5cm]{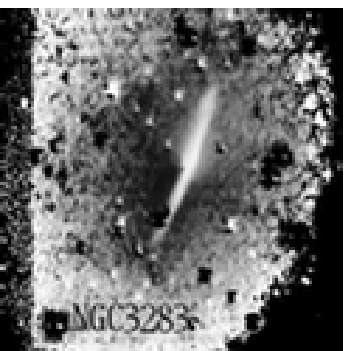}}\vspace{-6mm}
\subfigure{\includegraphics[width=5cm]{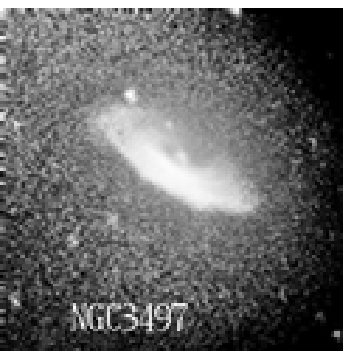}}
\subfigure{\includegraphics[width=5cm]{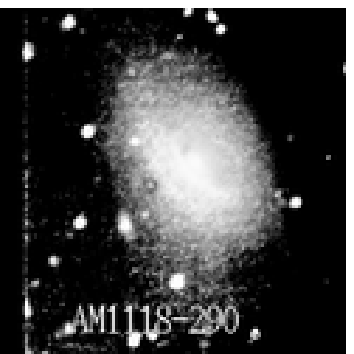}}\\
\subfigure{\includegraphics[width=5cm]{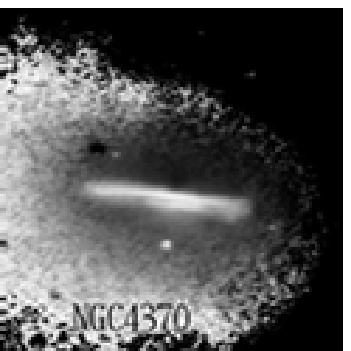}}\vspace{-6mm}
\subfigure{\includegraphics[width=5cm]{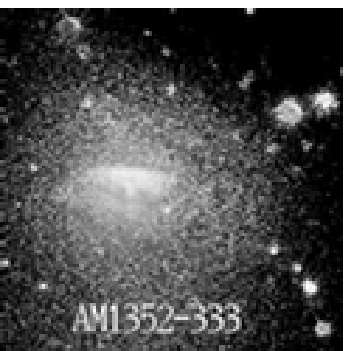}}
\subfigure{\includegraphics[width=5cm]{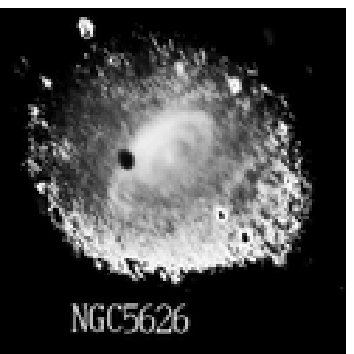}}\\
\subfigure{\includegraphics[width=5cm]{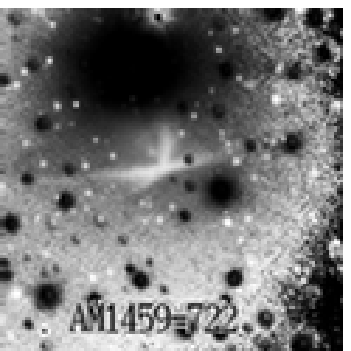}}\vspace{-6mm}
\subfigure{\includegraphics[width=5cm]{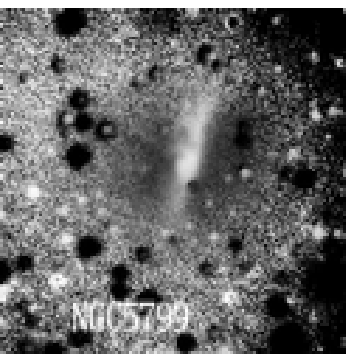}}
\subfigure{\includegraphics[width=5cm]{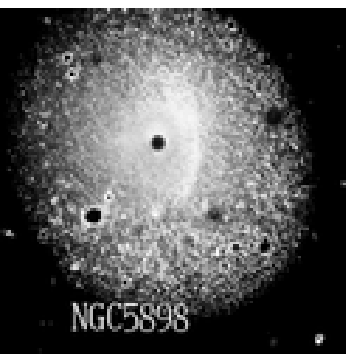}}
\\
\vspace{30pt}\caption{B-I colour-index maps of the sample galaxies. Brighter regions near the images center represent the suspected dust lanes.}
\label{fig:BImaps}
\end{minipage}
\end{figure*}

As is apparent from Fig.\ \ref{fig:BImaps}, the spatial distribution of the dust is clearly not that of a foreground screen between us and the light sources. We may use a representative value of $p=0.15$, estimated by Knapen et al.\ (1991) for the Sombrero galaxy, as a simple test case to estimate the effect of a dust lane located between us and the light source on the derived extinction law. Fitting a synthetic curve with $p=0.15$, illustrated in Fig.\ \ref{fig:EvsA}, with a linear trend shows a very significant deviation from the S\&M relation. However, an attempt to fit the sample galaxies under the assumption of $p=0.15$ in order to obtain the optical depth in each dusty location does not show a significant deviation from the results obtained with $p=0$, with the $R_\lambda$ values  only slightly decreasing. This may be explained if the optical depth in the dusty locations is relatively thin, i.e., $\tau_V<1$ as typical for our sample galaxies, where the deviation from the ``real'' linear relation may not be significant and within the error bars. 
We also note that due to the spatial resolution, only about a dozen dusty locations were measured within each galaxy, while it is possible that the dust geometry and pattern may change throughout the galaxy. Therefore, better spatial resolution is required in order to trace the possible variations of the dust lane location within the galaxy. 
One may reproduce the ``real'' linear relation from the observed one by fitting for different $p$ values. However, considering a more complex dust geometry, this task would require the use of more complex radiative transfer models.

We should also consider the possibility that part of the light from the dust lanes is due to the forward-scattering of light from stars in the galaxy. It is well-known that the amount of scattered light increases strongly towards shorter wavelength where the dust albedo is higher. Therefore, the scattering of light artificially increases the observed $R_\lambda$ and subsequently ``flattens'' the extinction curve. Bruzual, Magris \& Calvet (1988) have treated the complex case of multiple scattering in disk galaxies and demonstrated that, for an almost edge-on galaxy, $R_V$ can reach values as high as 5 for a reasonable amount of dust (also see Goudfrooij \& de Jong 1995). Emsellem (1995) suggested that the scattering of light by dust particles plays a crucial role on the observed extinctions of the highly inclined Sombrero galaxy; this is by acting as an additional source of radiation and thus reducing the observed attenuation by a factor of 2. He concluded that the relatively high extinction values of the Sombrero galaxy are due to the saturation of the observed attenuation for $A_V > 3$ mag.
Using Monte Carlo techniques for radiative-transfer calculations, Witt, Thronson \& Capuano (1992) and Baes, Dejonghe \& Davies (2005) showed that the effects of scattering can be important, even for small optical depths, and that neglecting or approximating the effects may lead to serious extinction errors. These authors also showed that the forward scattering strongly depends on the geometry of the dust and star clouds. For a uniform sphere of dust within a larger uniform stellar distribution, Witt et al.\ (1992) showed that only a few percent of the unextinguished light is forward-scattered for a total extinction typical for our sample galaxies (i.e., $\sim$0.5\%). Since the measured extinction values are not significally larger than the MW extinction values, and the extinction curves do not show any flattening towards the near-UV, it is fair to assume that the effect of scattering is negligible compared to the effect of the grain size, at least for those galaxies which are not highly inclined.

\begin{center}
\begin{figure*}
\begin{minipage}{150mm}
\includegraphics[]{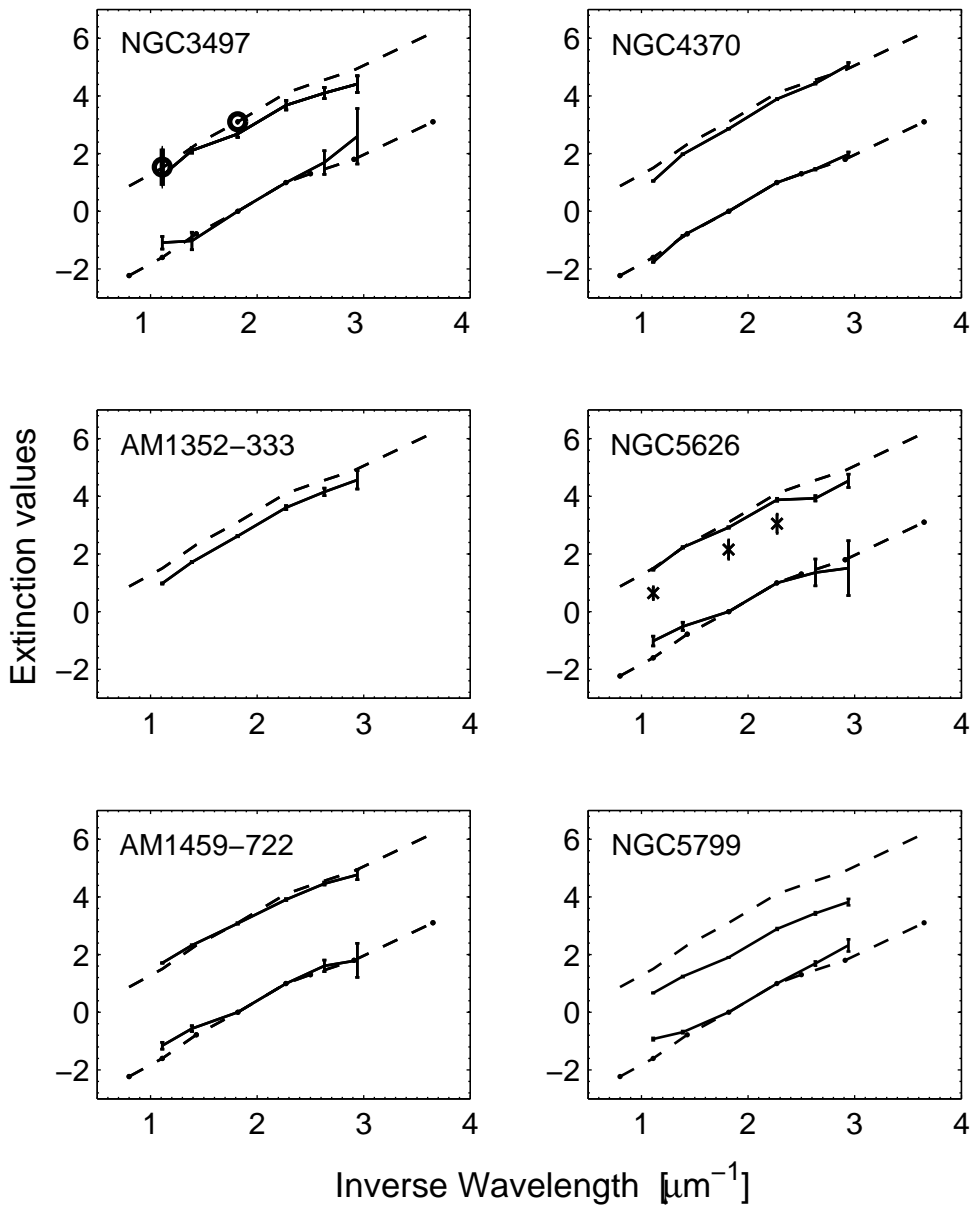} 
\caption{Extinction curves for the program galaxies (solid lines) along with the canonical curve for the galaxy (dashed lines) for comparison. The upper curves for each galaxy represent $R_\lambda$ values whereas the lower ones represent $\dfrac{A_\lambda-A_V}{A_B-A_V}$ values. Extinction values derived from previous papers are from Goudfrooij et al.\ (1994) and Patil et al.\ (2007), and are represented by crosses and open circles respectively. The error bars are $1\sigma$ errors. 
\label{fig:curves}}
\end{minipage}
\end{figure*}
\end{center}

\begin{center}
\begin{figure*}
\begin{minipage}{150mm}
\includegraphics[]{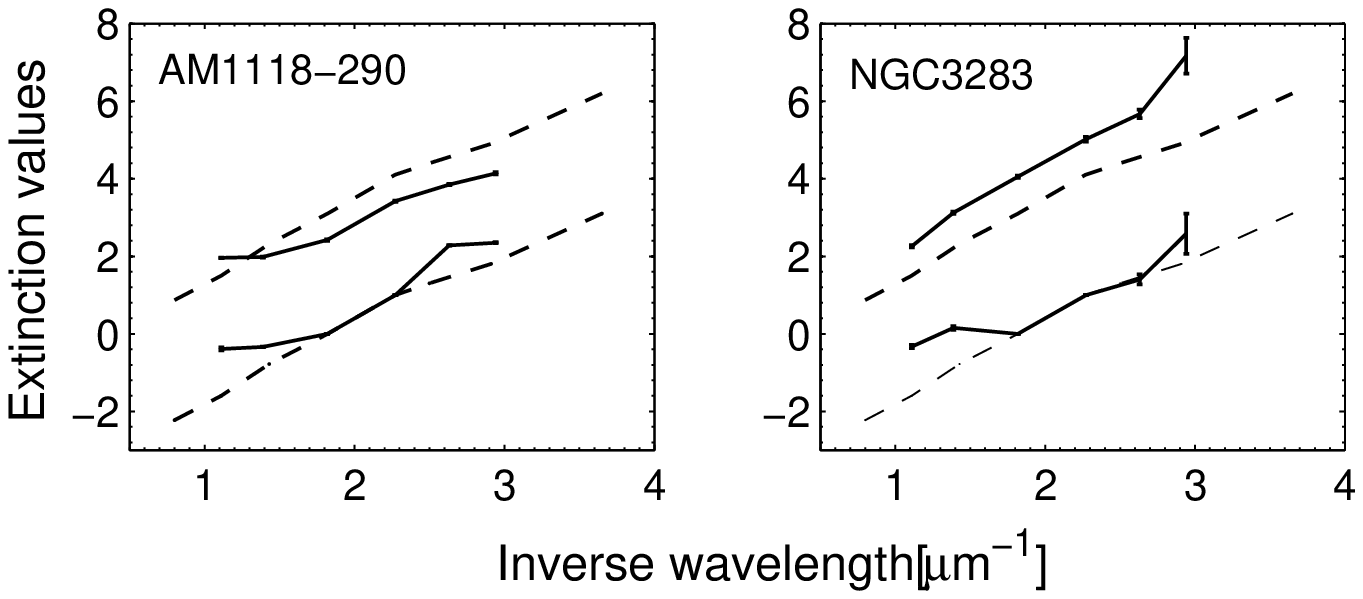} 
\caption{Extinction curves for NGC3283 and AM1118-290 (solid lines) along with the canonical curve for the galaxy (dashed lines) for comparison. These galaxies show noisy extinction plots, therefore the derivation of the extinction values is suspect. The upper curves for each galaxy represent $R_\lambda$ values whereas the lower ones represent $\dfrac{A_\lambda-A_V}{A_B-A_V}$ values. The error bars are $1\sigma$ errors. 
\label{fig:twocurves}}
\end{minipage}
\end{figure*}
\end{center}

We derived extinction curves and extinction values for eight of our nine sample galaxies, with the exception of NGC5898.
The obtained $R_\lambda$ extinction values are listed in Table \ref{t:Rvalues}, along with the canonical extinction values for the B, V, R \& I bands taken from S\&M for comparison. The extinction curves are plotted in Fig.\ \ref{fig:curves} and \ref{fig:twocurves}. For the case of the faint arc-like dust lane of NGC5898 (see B-I colour index map in Fig.\ \ref{fig:BImaps}), the measured V-band extinction per resolution element is rather small with values up to 0.1 mag. In addition, these galaxy images are relatively faint due to short exposure times (see Table  \ref{t:exposure}) and therefore the extinction is poorly determined.
For the case of NGC3283 and AM1118-290, the B-I colour-index maps show prominent dust lanes along the galaxies major axes. These galaxies show noisy extinction plots in all bands, thus extinction values are difficult to determine using only a linear regression fit, as has been suggested before. 

For the case of major-axis dust lanes it is possible that the dust lanes are actually dust disks seen edge on, therefore we should not expect the relation $A_\lambda=1.086 \, \tau_\lambda$ to remain valid. A 2D decomposition analysis of AM1118-290 shows a stellar disk that may be responsible for significant forward scattering from the major-axis dust lane. While NGC3283 shows a simple de Vaucouleurs profile with a boxy structure, as also indicated by Bertola et al.\ (1988), the same authors observed radio jets from the galaxy center, implying the presence of an active central source that could produce a significant forward-scattering from the major-axis dust lane. 

We should note that the derived $R_V$ values for NGC3283 and AM1118-290 are $\sim$2.42 and $\sim$4.05, respectively. The first may imply that forward scattering, which tends to increase the extinction values, may not be dominant in the case of NGC3283. The extinction curves for these two galaxies are plotted in Fig.\ \ref{fig:twocurves}. 
We also note that for the case of NGC4370, another major-axis dust lane elliptical, there were no signs of a luminous central body in our 2D decomposition analysis and although the $\dfrac{A_\lambda}{A_V}$ values measured in various dusty regions produce somewhat noisy plots, we obtain extinction values very similar to the Galactic extinction values. 
\begin{table*}
 \centering
 \begin{minipage}{140mm}
\caption{$R_{\lambda}$ values \label{t:Rvalues}}
\begin{tabular}{|l|cccccc|}
\hline
Object    & $R_{U1}$       & $R_{U2}$        & $R_B$            & $R_V$            & $R_R$            & $R_I$ \\
(1)       & (2)             & (3)              & (4)              & (5)              & (6)              & (7) \\
\hline
NGC3283    & $4.14 \pm 0.03$ & $ 3.85 \pm 0.02$ & $ 3.42 \pm 0.01$ & $ 2.42 \pm 0.01$ & $ 1.98 \pm 0.01$ & $ 1.96 \pm 0.01$\\
NGC3497    & $4.41 \pm 0.29$ & $ 4.10 \pm 0.19$ & $ 3.68 \pm 0.16$ & $ 2.68 \pm 0.11$ & $ 2.11 \pm 0.09$ & $ 1.23 \pm 0.05$\\
NGC4370    & $5.07 \pm 0.09$ & $ 4.44 \pm 0.03$ & $ 3.89 \pm 0.02$ & $ 2.86 \pm 0.01$ & $ 1.98 \pm 0.01$ & $ 1.05 \pm 0.01$\\
AM1352-333 & $4.57 \pm 0.32$ & $ 4.15 \pm 0.13$ & $ 3.61 \pm 0.07$ & $ 2.62 \pm 0.02$ & $ 1.72 \pm 0.02$ & $ 0.98 \pm 0.02$\\
AM1118-290 & $7.17 \pm 0.46$ & $ 5.67 \pm 0.11$ & $ 5.02 \pm 0.06$ & $ 4.05 \pm 0.03$ & $ 3.12 \pm 0.03$ & $ 2.26 \pm 0.03$\\
NGC5626    & $4.54 \pm 0.23$ & $ 3.93 \pm 0.09$ & $ 3.88 \pm 0.05$ & $ 2.92 \pm 0.02$ & $ 2.23 \pm 0.02$ & $ 1.46 \pm 0.02$\\
AM1459-722 & $4.77 \pm 0.17$ & $ 4.46 \pm 0.05$ & $ 3.91 \pm 0.03$ & $ 3.08 \pm 0.02$ & $ 2.32 \pm 0.02$ & $ 1.71 \pm 0.01$\\
NGC5799    & $3.82 \pm 0.11$ & $ 3.43 \pm 0.04$ & $ 2.89 \pm 0.03$ & $ 1.90 \pm 0.01$ & $ 1.24 \pm 0.02$ & $ 0.67 \pm 0.04$\\ 
\hline 
MW galaxy  &    4.9          &   4.56           & 4.10             &3.10              & 2.32             & 1.5\\
\hline
\end{tabular}
\end{minipage}
\end{table*}
\subsection{An alternative method - comparing colour-index maps}
\label{S:indexcolourmaps}
We suggest here an alternative method for deriving extinction values, based on the comparison of colour-index maps of dusty early-type galaxies. As the traditional method described earlier, this alternative method also relies on the assumption of smoothly varying nature of the light distribution in early-type galaxies, in which the dust lanes or dust patches are only local disturbances. Considering the de Vaucouleurs and S\'{e}rsic profiles (eq.\ \ref{eq:devaucouleurs} and eq.\ \ref{eq:disk}) for early-type galaxies, one may assume that the brightness variation in different locations in the galaxy with respect to its central brightness is wavelength-independent, i.e., is similar for all bands. As a result, all colour-index maps of such a smooth early-type galaxy should exhibit constant values for the entire galaxy. The hidden assumption here is that the stellar population mix does not vary much across the galaxy; only the column density of stars presented to the line of sight changes with galacto-centric distance. This is justified by the shallow colour gradients measured for normal early-type galaxies (e.g., La Barbera et al.\ 2008) where the small colour gradients are explained as a metallicity effect.

If dust is present, and if this dust produces wavelength-dependent extinction in the optical, a colour-index map, e.g., a B-V map, is expected to redden at the dusty locations with respect to the constant colour of the dust free regions. This reddening may be linked to the difference in extinctions $A_B-A_V$. Therefore, by comparing different colour values of dusty regions one may obtain certain  extinction values and study the dust properties.

We compared the various $\lambda$-V maps with B-V maps for each galaxy by measuring the colour values in different regions in the galaxies. As expected, the maps show constant colour values throughout the galaxies and redden where the dark lanes appear. As before, we fit a linear regression between the $\lambda$-V and the B-V values, where the slope represents the extinction value $\frac{\mbox{E($\lambda$-V)}}{\mbox{E(B-V)}}$ and the intercept with the $\lambda$-V axis depends on the constant colour of the dust-free galaxy in the different bands. This means that if one can measure these colour values in different dusty regions, then one does not need to specifically measure the constant colours, but only to assume that such values do exist.

This method of deriving extinction values has the disadvantage of being somewhat less informative in comparison with the traditional method outlined in \S~\ref{S:extinctionval}. Using this alternative method one cannot measure the extinction in each band nor to derive $R_\lambda$ values. The latter may be of importance if one wishes to examine whether the target galaxy's extinction curve lies above or below the MW extinction curve (see also \S~\ref{txt:interp}). Nevertheless, comparing the $\frac{\mbox{E($\lambda$-V)}}{\mbox{E(B-V)}}$ extinction values of different galaxies with the Galactic extinction values we may examine whether these extinction curves run parallel to the Galactic extinction curve or whether they deviate from this linear trend, implying different extinction laws.
Despite its pitfalls, the alternative method has the advantage of being very easy-to-use and straightforward, without requiring any model fitting but only assuming that such a smooth surface brightness distribution and colour model do exist. Moreover, while the traditional method relies on the dust-free regions of the galaxy in order to model the entire underlying galaxy, the alternative method may be used without such requirements, relying solely on the dusty regions themselves. This means that the alternative method may be used even in cases where the dust obscures large fractions of the underlying galaxy, or in cases where the modeling of the underlying galaxy is not simple. 
In addition, the alternative method may be favoured when the image resolution is poor and there are not sufficient datapoints to be fitted by a linear regression. In this case, one may have to derive the colour of the galaxy by measuring dust-free regions.
We plotted the extinction values obtained using the alternative method along with the extinction values obtained using the traditional method in Fig.\ \ref{fig:curves}. As these plots show, the alternative method produces extinction values with larger errors compared with those produced using the traditional method.

\section{Interpretation of dust properties}
\label{txt:interp}
We adopt the Mathis, Rumpl \& Nordsieck (1977) dust grain model succesfully used in reproducing the MW extinction curve in the optical-UV part of the spectrum. By varying the grain size parameters and the abundance ratio of graphite and silicates we reproduce the galactic extinction curve and determine the dust grain characteristic size. We find the average characteristic size in our sample galaxies to be similar to the characteristic MW dust grain size. We also estimate the dust mass in the sample galaxies from optical extinction measurements and find them to be significantly smaller than the dust mass derived using far-IR data.
\subsection{The derivation of an extinction law}
Assuming that the chemical composition of the extragalactic dust grains is uniform throughout each galaxy and is similar to that of the dust in our Galaxy, one can compute the relative grain size and the total dust mass in the sample galaxies by using available models for the chemical composition and the shape of the dust grains. 
The term ``interstellar dust'' refers to different materials with rather different properties and there are significant differences in extinction between low-density diffuse dust and dust in the inner and outer parts of molecular clouds. Therefore, for the case of dust lane galaxies, we can only refer to a ``mean'' or typical extinction law, assumed to represent the large-scale diffuse dust.

Since the total extinction linearly depends on the extinction cross-section, we may express $R_\lambda$ as:
\begin{equation}
R_\lambda=\dfrac{C_{ext}\left( \lambda\right) }{C_{ext,B}-C_{ext,V}}.
\label{eq:R_V}
\end{equation}
where $C_{ext}$ represents the extinction cross-section of grains in a volume unit.
The mean cross-section for spherical particles with a size distribution of $n_i\left( a\right) da$, where $i$ indicates a certain component and $a$ represents a certain grain size, can be written as:
\begin{equation}
\left\langle C_{ext}\left( \lambda\right)\right\rangle  = \left\langle Q_{ext}\left(a,\lambda \right) \pi a^2 \right\rangle =\dfrac{\Sigma_i \int_{a_-}^{a_+}Q_{ext}\left( a,\lambda\right)\pi a^2 n_i\left( a\right)da}{\Sigma_i \int_{a_-}^{a_+}n_i\left(a \right)da }
\label{eq:cross-section}
\end{equation}
where $a_-$ and $a_+$ represent the lower and upper cutoffs of the size distribution, respectively; $\lambda$ is the wavelength; $Q_{ext}\left(a,\lambda \right)$ is the ratio of the extinction cross-section to the geometrical cross-section and is defined as the extinction efficiency.

Several grain models, invoking different size distribution functions and mixtures of chemical compositions, are proposed in the literature. Since we restrict ourselves to the extinction in the optical part of the spectrum, it is possible to use a simple  model that can account for the MW optical extinction curve. We adopt the Mathis, Rumpl \& Nordsieck (1977; hereinafter MRN) two-component model consisting of individual spherical silicate and graphite grains with an adequate mixture of sizes. 
The MRN model assumes uncoated refractory particles having a power-law size distribution of
\begin{equation}
n(a)\,da=n_H A_i \, a^{-3.5}\,da ,
\label{eq:Mathis77}
\end{equation}
where $a$ represents the grain size, $n_H$ is the hydrogen number density and $A_i$ is the abundance of component $i$. The size distribution is truncated at the upper end at $a_+=0.25\mu$m, with the lower end of sizes extending downward to $a_-=0.005\mu$m.
The MRN model was also adopted by Draine \& Lee (1984; hereinafter DL84) who succesfully used the Mie scattering theory for homogeneous spherical particles in order to reproduce the MW extinction curve in the optical-UV part of the spectrum.
DL84 adopt a graphite-to-silicate abundance ratio of  $\dfrac{A_{g}} {A_{s}} \simeq 1.12$ to optimize the fit of their synthetic extinction curve to the observational data. We emphasize that this model consists only of spherical particles, while observations of interstellar polariztion indicate that an appreciable fraction of the grains must be nonspherical. Since we do not deal here with polarization, we disregard non-spherical grains.

Using eq.\ (\ref{eq:R_V}) and eq.\ (\ref{eq:cross-section}), one can calculate the theoretical extinction values, where the extinction efficiency for specific wavelength, grain size and refractive index is calculated using the standard Mie theory (Mie 1908; Debye 1909) given the dielectric functions for graphite and `astronomical silicate' derived by DL84 and published in tabular form (Draine 1985; see also http://www.astro.princeton.edu/$\sim$draine/dust/dust.diel.html). The method for calculating the extinction values is also outlined by Steenman \& Th\'{e} (1989).
\subsection{Dust characteristic size}
We now estimate the characteristic particle size relative to that responsible for the Galactic extinction law. 
This was done in previous papers (Goudfrooij et al.\ 1994 and Patil et al.\ 2007) by comparing the wavelength-dependence of the extinction efficiency of spherical dielectric grains with that of the observed $R_\lambda$ extinction values.
This method relies on observations indicating that the R$_{\lambda}$ curves usually vary approximately linearly with inverse wavelength, as observed for the Galactic extinction curve and for extinction curves of different galaxies as derived by Goudfrooij et al.\ 1994 and Patil et al.\ 2007. Therefore, the different extinction curves seem to run along or parallel to the Galactic curve.
Considering the optical part of the spectrum, and the upper and lower cutoff sizes estimated for the grains, Goudfrooij et al.\ (1994) assumed $Q_{ext} \propto \dfrac{a}{\lambda}$ (see also van de Hulst 1957). However, considering eq.\ (\ref{eq:R_V}) and eq.\ (\ref{eq:cross-section}), it seems that this assumption leads to a situation where $R_\lambda$ is grain size independent.
Therefore, irrespective of the grain model used, all $R_\lambda$ values remain constant and higher order terms of the series expansion of the extinction efficiency are necessary in order to estimate the characteristic particle size relative to that responsible for the Galactic extinction curve. 

Adopting the MRN grain model, different extinction curves can be reproduced by evaluating the extinction efficiency under the assumption of spherical grains and using the available Mie theory code derived from BHMIE (Bohren \& Hoffman 1983)\footnote{The code was written by Bruce T. Draine and is available at \textbf{ftp://ftp.astro.princeton.edu/draine}}. 
This fitting scheme should include several free parameters such as the size distribution power law, the abundance  and the upper and lower size cutoff values ($a_+$ and $a_-$) for both graphite and silicate grains (see also Mathis \& Wallenhorst 1981;  Steenman and Th\'{e} 1989, 1991). 

Considering the extinction efficiency calculations and the different extinction curves, which we find to run parallel to the Galactic extinction curve, we can generally conclude that increasing (decreasing) the number of upper end size grains with respect to the Galactic grain population produces larger (smaller) $R_\lambda$ values. Such an increase of the average grain size may be the consequence of a flatter size distribution power law, or of an increase in $a_+$ and $a_-$. 
Therefore, we conclude that extinction curves that lie above (below) the Galactic extinction curve correspond to larger (smaller) characteristic grain sizes with respect to the typical Galactic grain size.
However, we note that varying the abundance ratio of the two silicate-graphite components may also account for different extinction values. Calculations show that increasing the $\dfrac{A_s}{A_g}$ ratio, as well as reducing the characteristic grain size, leads to a decrease of the extinction values and vice versa.

We note that our measured extinction values in the near-IR to near-UV region cannot properly constraint all the free parameters mentioned above. Therefore, a precise fit should include a better determined extinction curve and an extension of this extinction curve to the UV region. 
Since the size distribution power law suggested by the MRN grain model appears to be consistent with the power law expected from grain-grain collisions (Biermann \& Harwit 1980), and since the observed silica and carbon abundances in the ISM are approximatelly equal, we consider here only a variation of $a_+$. As far as $a_-$ is concerned, surface effects and quantum mechanics effects make Mie calculations suspect for particles smaller than about 0.01 $\mu$m. 

Using eq.\ (\ref{eq:R_V}) and the extinction cross-section calculations we obtain the upper size cutoff value for the graphite-silicate grains mixture from the best fits to the extinction values for each sample galaxy and for the Galaxy. 
We find the average upper size cutoff value in our sample galaxies (see Table \ref{t:avalues}) to correspond only very approximately with the value of 0.22 $\mu$m obtained for the MW. The $a_+$ values are mostly lower than 0.22 $\mu$m, with only one (AM1118-290) significantly higher. 

\subsection{Dust mass estimation}
In order to derive the dust mass we use the dust column density from
\begin{equation}
\Sigma_d=l_d{\times}\int\limits_{a_-}^{a_+}\frac{4}{3}{\pi}a^3{\rho}_dn\left(a\right)da 
\end{equation}
where ${\rho}_d$ gives the specific grain density and $l_d$ represents the dust column length along the line of sight. The value of $l_d \times n_H$ can be inferred from the measured total extinction and the calculated extinction efficiency in the V band using eq.\ (\ref{eq:cross-section}) and $a_+$ and $a_-$ obtained for each sample galaxy. The specific grain densities of graphite and olivine-like `astronomical silicate' grains are conservatively estimated as 2.26 gr cm$^{-3}$ and 3.3 gr cm$^{-3}$, respectively (DL84).
The total mass is derived by integrating the dust column density over the image areas occupied by dust lanes and is therefore given by $M_d={\Sigma}_d{\times} \mbox{Area}$, where $M_d$ is expressed in solar mass units.
We also note that it is also possible to derive the optical dust mass using the alternative method described in \S~\ref{S:indexcolourmaps}, i.e., using the differences in the extinction and cross-sections in two optical bands. However, the latter requires subtracting the constant dust-free galaxy colours from the colour values where dust is present in order to obtain the correct difference of extinctions.

The dust mass can be estimated independently from the far-infrared emission of each galaxy by using the dust grain temperature calculated from the IRAS flux densities at 60{$\mu$}m and 100{$\mu$}m using $T_d=\left( \frac{S_{60}}{S_{100}}\right)^{0.4}$ (Young et al.\ 1989). The dust mass is then computed for each sample galaxy using the relation (Hildebrand 1983)
\begin{equation}
M_d=\frac{4}{3} a \rho _d D^2 \frac{F_{\nu}}{Q_{\nu}B_{\nu}(T_d)}
\end{equation}
where $a$, ${\rho}_d$ and $D$ are the grain radius, specific grain mass density and distance of the galaxy in Mpc (assuming H$_0$=70 km s$^{-1}$ Mpc$^{-1}$), respectively; $F_{\nu}$, $Q_{\nu}$ and $B_{\nu}(T_d)$ are the observed flux density, grain emissivity and the Planck function for the temperature $T_d$ at frequency $\nu$, respectively. 
We can evaluate the graphite-silicate mixture mean emissivity by using published extinction efficiencies values and assuming thermal equilibrium. Since the quantity $a/Q_{\nu}$ is independent of $a$ for  $a \gg \lambda$ (Hilderbrand 1983), we calculate $Q_{\nu}$ for a mixture of $0.1 \mu$m graphite and silicate grains at $\lambda=100 \mu$m using the tabulated data from Draine (1985).
These dust mass estimates represent lower limits, since IRAS was insensitive to dust cooler than about 20K which emits mostly at wavelengths longer than $100 \mu$m.
Table \ref{t:Mass} lists the estimated dust mass from the total optical extinction and the estimated dust mass based on IRAS flux densities taken from the catalog of Knapp et al.\ (1989) for bright early-type galaxies.

\begin{table*}
 \centering
 \begin{minipage}{140mm}
\caption{Dust properties \label{t:avalues}}
\begin{tabular}{|lccccc|}
\hline
Object     & $a_{max}(\mu\mbox{m}) $ & \multicolumn{2}{c}{IRAS flux (mJy)} & $ \mbox{Log}\left( \frac{M}{M_ \odot}\right) _{d,optical} $ & $ \mbox{Log} \left(\frac{M}{M_ \odot}\right)_{d,\mbox{IRAS}}$\\
   {}      &  {}              & 60$\mu$m     & 100$\mu$m      &     {}           &    {}        \\
(1)        &      (2)         &   (3)        &  (4)           &     (5)          &   (6)        \\
\hline
NGC3283    & $0.19 \pm 0.01 $ & $1200 \pm 68$& $4440 \pm 207$ & $5.67 \pm 0.03$  & $6.48\pm0.08$ \\ 
NGC3497    & $0.18 \pm 0.01 $ & $278$        & $<1347$        & $6.03 \pm 0.03$  & $<6.43$      \\ 
AM1118-290 & $0.30 \pm 0.01 $ & $190 \pm 35$ & $800 \pm 128$  & $6.26 \pm 0.02$  & $6.76\pm0.25$\\
NGC4370    & $0.18 \pm 0.01 $ & $990 \pm 40$ & $2900 \pm 77$  & $4.46 \pm 0.03$  & $4.96\pm0.05$\\ 
AM1352-333 & $0.16 \pm 0.01 $ & $110 \pm 35$ & $130 \pm 365$  & $5.84 \pm 0.04$  & $4.54\pm2.83$\\ 
NGC5626    & $0.20 \pm 0.01 $ & $210 \pm 23$ & $760 \pm 69$   & $5.93 \pm 0.03$  & $6.44\pm0.15$\\ 
AM1459-722 & $0.22 \pm 0.01 $ &              &                & $5.96 \pm 0.03$  &              \\
NGC5799    & $0.10 \pm 0.01 $ & $320 \pm 21$ & $2370 \pm 294$ & $5.17 \pm 0.07$  & $6.96\pm0.21$\\ 
\hline
\end{tabular}
\label{t:Mass}
\end{minipage}
\end{table*}
\section{Discussion}
\label{txt:disc}

We have studied the extragalactic dust extinction in nine dust-lane early-type galaxies. 
The $R_V$ values derived above are close to the standard Galactic value, at least for most of the derived extinction curves. The mean value for these galaxies, $R_V=2.82\pm0.38$, corresponds to the value derived by Patil et al.\ (2007) for galaxies with well-settled dust lanes, i.e., 2.80, supporting an explanation that the characteristic grain size responsible for the optical extinction in such galaxies is slightly smaller than that in the MW, but within the error bars. These $R_V$ values, with the exception of AM1118-290, are also in the range of values derived by Goudfooij et al.\ (1994), i.e., 2.1 to 3.3, for which they estimated a characteristic dust grain size up to 30\% smaller compared to standard Galactic dust grains. 
We note that the extinction value derived for NGC5799 has an exceptionally low value of $R_V=1.90 \pm 0.01$; this is similar to the lowest extinction value obtained by Patil et al.\ (2007)  and Goudfrooij et al.\ (1994). 

As described in \S~\ref{S:extinctionval}, the effects produced by some foreground light cannot account for lower $R_V$ values and, if it all, dust uniformly embedded with stars or a dust lane in a distinct location within the galaxy will have the effect of producing larger values of $R_V$ for the observed extinction law. 
Another potential pitfall in the derivation of the extinction values is the possibility of star formation in the dust lanes. Goudfrooij et al.\ (1994) discussed this effect thoroughly and concluded that, due to the light emitted from young massive stars, one may underestimate the extinction, especially at the shorter wavelengths. Thus, the real extinction curve may be somewhat more concave than the measured one and with larger extinction values. However, this may not account for the relatively low extinction values of NGC5799, for which Brosch et al.\ (1990) concluded from the relatively low ratio between the blue and 60$\mu$m flux densities that it does not form stars at a very high rate in its ISM torus, and therefore star formation should have a negligible effect on its derived extinction values. This may also be concluded for the rest of the sample galaxy.
For most galaxies, we may also neglect the effect of forward scattering in dusty regions on the observed extinction curve. As already discussed in \S~\ref{S:extinctionval}, this effect leads to apparently lower extinction values towards the blue, making the extinction curve flatter than the Galactic curve, whereas such flattening is not evident in any of our sample galaxies. 

We obtained extinction curves for eight of our nine sample galaxies, with the exception of NGC5898. Concerning the latter, for which Patil et al.\ (2007) obtained $R_V=3.15\pm0.46$, the difficulty in determining its extinction curve may lie in the short SALT exposure times. 
We also found that NGC3283 and AM1118-290 show noisy $\dfrac{A_\lambda}{A_V}$ plots in most bands, suggesting a possibly complex dust structure or the presence of significant forward scattering. The latter seems more reasonable for AM1118-290 which produces larger $R_\lambda$ values with respect to the Galactic extinction values. 
For each of the NGC5799 and AM1459-722 galaxies Brosch et al.\ (1989) derived a flat extinction curve in the B \& V filters, which did not correspond with the Galactic extinction trend or values. It is therefore reassuring to derive extinction curves which run parallel to the Galactic extinction curve, though we find NGC5799 to have significally smaller grain sizes compared with the Galactic dust. The NGC5626 extinction law derived by Goudfooij et al.\ (1994), with $R_V=2.15$, implies significally smaller grains in comparison with $R_V=2.98$ we found, the latter being closer to the Galactic value.
The extinction values for NGC3479 are smaller than those derived by Patil et al.\ (2007) in the V and I bands, but still within the error bars. 
The major-axis dust lane elliptical NGC4370 shows an extinction curve very similar to that of the Galaxy. While the effect of forward scattering may be important in the case of major-axis dust lane disk galaxies, it does not seem significant for the case of pure ellipticals such as NGC4370. Such a possible effect, as well as that of a more complex dust structure, might reduce the observed total extinction of NGC4370 with respect to its optical depth. We did not find previously determined extinction values for the NGC3283, AM1118-290, NGC4370 and AM1352-333 galaxies with which to compare our results.

We introduced an alternative method to measure the extinction values $\frac{\mbox{E($\lambda$-V)}}{\mbox{E(B-V)}}$ by comparing colour-index maps. Although less informative, this method does not require fitting a model for the underlying galaxy, thus its results are not biased by the fitting procedures and may be compared with the traditional method. The comparison shows that the $\frac{\mbox{E($\lambda$-V)}}{\mbox{E(B-V)}}$ values are generally consistent with the Galactic extinction law and with the trends of the matching $R_\lambda$ curves. An exception is NGC3283, which shows a similar deviation from the Galactic extinction law in both extinction curves.

Although we started this study with the intention of finding differences between the properties of extragalactic dust in our objects and the properties of MW dust, we are compelled by the results to conclude that no major differences were found. It is possible that such differences, if they exist at all, may be detected in the space-UV.

Using the optical extinction values we also evaluated the dust mass for each of the sample galaxy, and found it to lie in the range $\sim 10^4$ to $10^7M_{\odot}$. This is in good agreement with the earlier estimates for early-type galaxies with dark lanes (Brosch et al.\ 1990; Goudfooij et al.\ 1994; Sahu et al.\ 1998; Ferrari et al.\ 1999; Dewangan et al.\ 1999; Tran et al.\ 2001; Patil et al.\ 2002 and Patil et al.\ 2007). The use of the optical method to derive dust masses is based on the assumption that the dust forms a foreground screen for the galaxy. This means that we do not consider the dust to intermix with stars within the galaxy, nor the possible twisting of dark lanes around or through the host galaxies. Therefore, this method provides only a lower limit to the true dust content of the host galaxies and it should be compared with studies using other methods. 
Comparison of dust masses derived using optical extinction with those derived using IRAS flux densities reveals that the latter are up to an order of magnitude higher than the first. An exception is AM1352-333 where the dust mass obtained using far-IR data is poorly determined.
Our results are in good agreement with previous estimates (Goudfooij \& de Jong 1995; Sahu et al.\ 1998; Dewangan et al.\ 1999; Patil et al.\ 2007), which also concluded that dust masses derived using IRAS flux densities are usually larger by up to an order of magnitude than those estimated using the optical method, so that the discrepancy is more significant for ellipticals than for lanticulars. Moreover, since IRAS is not sensitive to cold dust, which emits the bulk of its radiation longwards of 100 $\mu$m, the IRAS dust mass estimates are a lower limit for the true dust mass which may even be an order of magnitude higher (Temi et al.\ 2004).
Goudfrooij \& de Jong (1995) suggested that this discprepancy may be solved if the interstellar dust exists in early-type galaxies as a two-component medium, where the less massive component is optically visible in the form of a dust lane or dust patches, and the more massive component is the diffuse dust which is distributed over the galaxy.

\section{Conclusions}
\label{txt:summ}

We presented SALT observations using the SALTICAM of early-type
galaxies with dust lanes obtained in order to derive the
extragalactic extinction law down to the ultraviolet atmospheric
cutoff. Our work also extends that done by previous authors since our derived extinction laws
reach to $\lambda^{-1}\approx2.94$. This extension demonstrates the necessity of performing such studies at as short a wavelength as possible.

Extinction curves derived for the sample galaxies run parallel to the canonical MW curve, implying similiar properties between MW canonical grains and dust in the extragalactic enviroment. 
We derived the ratio between the total V band extinction and the selective B and V extinction $R_V$ for each galaxy,  and obtained an avergage of $2.82\pm0.38$. These results are in agreement with Patil et al.\ (2007), indicating that galaxies with well-defind dust lanes tend to have slightly smaller $R_V$ values with respect to the Galactic value, but within the error bars. This suggests that the characteristic grain sizes responsible for the optical extinction are similar in size to the canonical MW grains.
We verified our results by using an alternative method to derive the extinction values, which is based on comparing colour-index maps. 

The dust content derived by the optical extinction method for the sample galaxies is in the range $\sim 10^4$ to $10^7M_{\odot}$; this also is in agreement with Patil et al.\ (2007) and with previous studies.

\subsection*{Acknowledgments}
The observations analyzed in this paper were partly obtained while NB was a sabbatical visitor at the
South African Astronomical Observatory in Cape Town; NB is
grateful for this opportunity offered by the SAAO management. We
are grateful for the generous allocation of SALT observing time
during the PV phase to complete this project.  We acknowledge the
use of products of the second Digitized Sky Survey produced at the
Space Telescope Science Institute under U.S. Government grant NAG
W-2166. The images are based on photographic data obtained using
the UK Schmidt Telescope. The UK Schmidt Telescope was operated by
the Royal Observatory Edinburgh, with funding from the UK Science
and Engineering Research Council (later the UK Particle Physics
and Astronomy Research Council), until 1988 June, and thereafter
by the Anglo-Australian Observatory. The blue plates of the
southern Sky Atlas and its Equatorial Extension (together known as
the SERC-J), as well as the Equatorial Red (ER), and the Second
Epoch [red] Survey (SES) were all taken with the UK Schmidt.

-------------------------


\newpage
\centering
\centering
\centering
\centering
\centering
\centering
\end{document}